**Tough Cortical Bone-Inspired Tubular Architected Cement-based Material**

*Shashank Gupta, Reza Moini\**

Department of Civil and Environmental Engineering, Princeton University, Princeton, NJ, USA

Email: Reza.Moini@princeton.edu

**Abstract.** Cortical bone is a tough biological material composed of tube-like osteons embedded in the organic matrix surrounded by weak interfaces known as cement lines. The cement lines provide a microstructurally preferable crack path, hence triggering in-plane crack deflection around osteons due to cement line-crack interaction. Here, inspired by this toughening mechanism and facilitated by a hybrid (3D-printing/casting) process, we engineer architected tubular cement-based materials with a new stepwise cracking toughening mechanism, that enabled a non-brittle fracture. Using experimental and theoretical approaches, we demonstrate the underlying competition between tube size and shape on the stress intensity factor from which engineering stepwise cracking can emerge. Two competing mechanisms, both positively and negatively affected by the growing tube size, arise to significantly enhance the overall fracture toughness by up to 5.6-fold compared to the monolithic brittle counterpart without sacrificing the specific strength. This is enabled by crack-tube interaction and engineering the tube size and shape, which leads to stepwise cracking and promotes rising R-curves. '*Disorder'* curves are proposed for the first time to quantitatively characterize the degree of disorder for describing the representation of architected arrangement of materials (using statistical mechanics parameters) in lieu of otherwise inadequate '*periodicity'* classification.

1. **Introduction**

Natural materials demonstrate an exceptional combination of two often competing mechanical properties, fracture toughness and strength, by assembling modest constituents into complex arrangements and hierarchical architectures.[1–9] Numerous biological materials, composed of large amounts of highly brittle minerals such as aragonite[10,11] and hydroxyapatite[12,13] achieve significantly higher fracture toughness by exploiting the purposeful arrangements, also known as materials' architecture, such as helical,[12–14] gradient,[15,16] layered,[17,18] suture,[19,20] and tubular.[21,22] High fracture toughness in these motifs is enabled by unique toughening mechanisms such as crack twisting in helical,[23] structural reorientation in gradient,[24] crack deflection along the interface in layered,[25] and interlocking in suture[26] architected materials.

On the contrary to natural counterparts, fracture toughness and strength, are often two competing mechanical properties in engineering materials.[27–29] For instance, technical and non-technical (e.g., concrete) ceramics can exhibit high strength but are limited by their low fracture toughness, relative to other classes of materials such as metals.[30] Concrete is the most commonly used human-made commodity in the world[31] and suffers from intrinsically low fracture toughness owing to the limited toughening mechanisms.[32–39] Consequently, this translates to the low fracture toughness under tension in cement-based materials including concrete,[32,33,36] mortar,[35,37,38] and cement paste.[34,37–39] More specifically, the low fracture toughness of the binding phase of the concrete matrix, the cement paste, is due to the limited ability to dissipate energy during the crack initiation and propagation throughout the microstructure.[40] Here, we present a viable mesoscale approach to engineer a toughening mechanism throughout the crack growth in the materials, by harnessing tubular voids that can promote crack-tube interaction and lead to stepwise cracking.

Tubular architecture has been exploited in several biological materials, such as human teeth,[41] human cortical bone,[42] horse hooves,[43] bamboo,[44] and ram horns,[45] to promote fracture resistance while serving metabolic functions.[25,46,47] The human femur bone, in particular, is composed of an outer cortical region[42] and an inner trabecular region[48] (**Figure 1**a). The cortical region consists of dense lamellated tubular osteons[3], whereas the inner trabecular region is porous and contains an interspersed and interconnected network of plates and rods[49] (Figure 1b). Cortical bone demonstrates exceptional fracture toughness due to its hierarchical structure that leads to resistance in crack initiation and propagation through various intrinsic and extrinsic multi-scale toughening mechanisms.[50] These mechanisms include molecular uncoiling and fibrillar sliding at the nano-scale,[50] micro-cracking and sacrificial bonds at the sub-micron scale,[50,51] as well as collagen-fiber bridging, uncracked-ligament bridging, crack deflection, and crack twisting at the micron scale.[3,7,50–52] Notably, we focus on crack deflection, triggered due to micro-scale osteons, also known as the Haversian system.[3,7,50–52] The osteons in the cortical bone are embedded in the organic matrix encapsulating Haversian canals for metabolic function and are surrounded by a particular weak interface known as cement lines[52] (Figure 1b, c). The cement lines have a specific mechanical characteristic as they have 10-fold lower tensile strength compared to the osteons they surround.[53] The typical elliptical cement lines, due to the weaker tensile strength, provide microstructurally preferred pathways for interaction with a propagating crack.[52] Hence, cement lines trigger unique toughening mechanisms in the cortical bone that involve in-plane crack deviation from a straight path and crack deflection around osteons, as schematically illustrated in Figure 1c[54,55]. This type of crack deflection prevents a brittle fracture [51] and is the key toughening mechanism investigated in this study to inform an initial design of tough engineering counterparts.

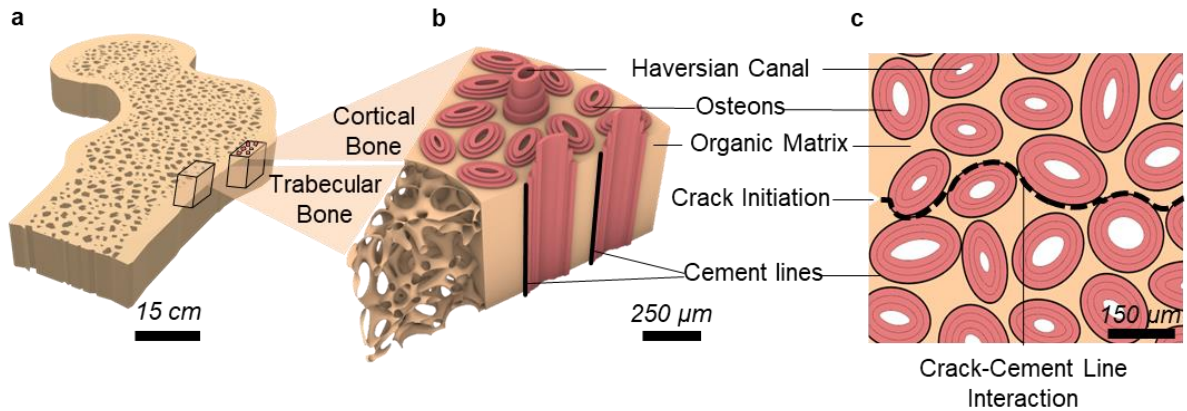

**Figure 1. Architecture of cortical bone.** a) Schematic cross-section of human femur bone illustrates the dense outer cortical bone and porous inner trabecular bone. b) Cortical bone architecture depicts the presence of weak cement lines surrounding tubular osteons, which leads to c) crack-cement line interaction, providing a path for in-plane crack-deflection from a straight brittle fracture as the toughening mechanism.

More specifically, here the particular interaction between the crack and the tubular geometry surrounding osteons found in cortical bone at the micro-scale is exploited for engineering a toughening mechanism in brittle paste material. This is enabled by circular and elliptical hollow tubular architecture at the meso-scale based on cortical bone.[37] More specifically, the design of the tubular architected materials is inspired by the consideration of the volume fraction[56] (45 - 65%) and shape (aspect ratio)[57] (nearly circular to elliptical, with an aspect ratio of $e$ = 1.02 - 2.62) of osteons (surrounded by cement line) in cortical bone. A periodic but *disordered* arrangement is considered in this study for the design of circular and elliptical tubes. The *disorder* is proposed as a more suitable metric, in contrast to periodicity (as a binary metric), to quantify the arrangement as a spectrum from a statistical mechanism perspective using radial distribution function, and two order parameters ($T$, $Q$). We hypothesize that engineering the crack-tube interaction toughening mechanism through designing tubular architected brittle cement-based material may lead to enhanced fracture toughness in an otherwise purely brittle material.[37]

While maintaining volume fraction and aspect ratio design parameters based on Haversian system, it should be noted that our bio-inspired design approach does not attempt to mimic the hierarchical arrangements of osteons observed in cortical bone. Rather, here we draw upon the one of the underlying toughening mechanisms of crack-cement line interaction as an inspiration to expand upon. This approach to bio-inspired design facilitates understanding and engineering a specific toughening mechanism, in contrast to the complexities that might arise from attempting a full emulation of the hierarchical structures of cortical bone.

To date, a few studies have only explored the design of composites inspired by cortical bone architecture by making use of inclusions (in the voids) to improve energy absorption (toughness),[58] fracture toughness,[59] and flexural strength.[59,60]. For instance, bone-inspired fibers-reinforced composites have been fabricated to enhance fracture toughness, by enabling crack deflection toward the inclusion boundaries, compared to solid laminated counterparts without inclusions. However, mere crack deflection is prevented at a small void size, and if present at a large void size, can be macroscopically undermined due to insignificant contribution to the stress required to advance the crack at the notch residing behind a void.[61] The bone-inspired composites exhibited limited improvement of up to 26% in fracture toughness and 30% in flexural strength compared to conventionally laminated counterparts[59,60] or 6.5 fold higher energy absorption (toughness) and 3.7 folds lower strength compared to solid counterparts by making use of stiff inclusions in a soft matrix.[58] In other studies, cast fiber-reinforced composites with tubular voids have been proposed to engineer auxeticity.[62,63] Another study investigates tubular voids with circular designs in ductile material to enhance the fracture toughness.[64] This research demonstrates the improvement in fracture toughness in ductile material for a notch with a blunt tip located *on* the tube that competes with the increasing circular tube size for higher fracture toughness. This approach couples the effect of the tube size on the notch tip (crack blunting) with the effect of the surrounding tubes on the (blunt) notch (hole-hole). In contrast, here we examine the effect of the circular and elliptical tube(s) *ahead* of the crack tip on a sharp notch, followed by the investigation of crack-tube interaction that enable stepwise cracking, where we hypothesize the increasing tube size impacts fracture toughness both negatively (due to increased stress intensity factor) and positively (due to enabling stepwise cracking). Several bio-inspired strategies observed at the micron scale in cortical bone have been adapted in human-made materials at milli-meter scale.[58–60]. Moreover, while limited studies have explored tubular geometry in composite and ductile material, the toughening mechanisms in brittle material counterparts have not been readily reported.

Contrary to the previous works where voids and inclusions are used to enhance crack deflection, we harness a new strategy to exploit the geometry (size, shape or aspect ratio) in tubular architecture itself as purposeful defects to engineer a new toughening mechanism and enhance fracture characteristics. Engineering the critical stress intensity factor (fracture toughness) through understanding the interactions of crack with the tube assists in allowing for a macroscopically effective stepwise cracking and highly non-brittle failure in an otherwise brittle material.

## 2. Result

To fabricate the tubular architected cement-based materials, we use a hybrid 3D-printing and casting approach. Firstly, a tubular template mold is additively manufactured using 3D-printing with polyvinyl alcohol (PVA) (**Figure 2**a). Furthermore, the negative of the template is fabricated by pouring two-part urethane rubber into the PVA template, which is subsequently dissolved to create a urethane silicon mold (Figure 2b). Tubular architected cement-based materials are cast (Figure 2c) into this silicon mold. Additional details of the fabrication process are described in the Methods.

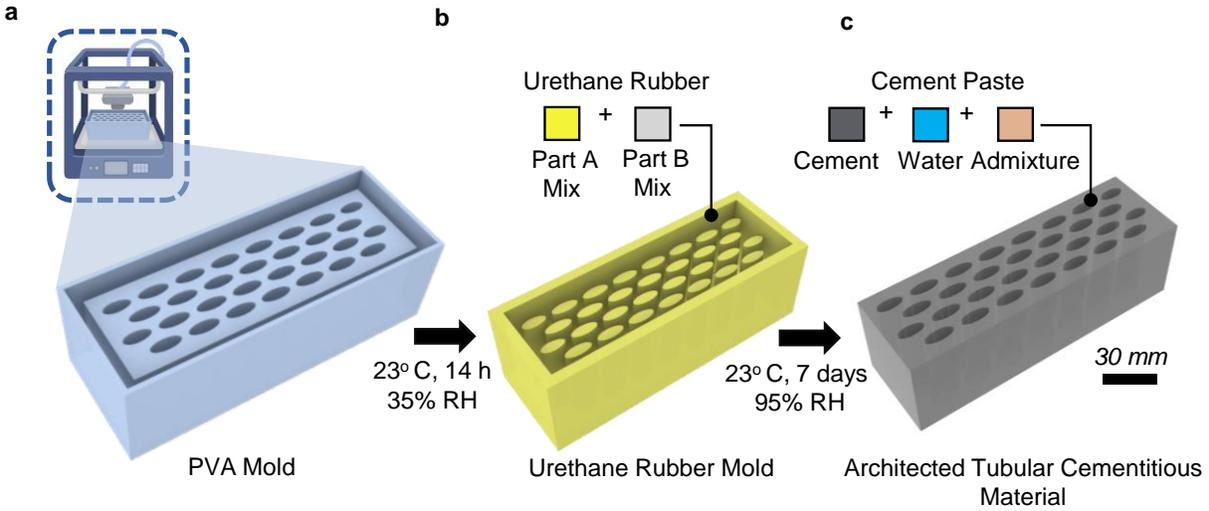

**Figure 2. Fabrication of tubular architected cement-based materials.** a) 3D-printing of a PVA template mold, followed by b) generation of a negative urethane rubber mold by dissolving the PVA template, and c) casting tubular architected cement-based material.

To study the effect of tube ellipticity on the mechanical response of the material, the aspect ratios ($e$) of the tubes vary from 1.0 to 3.0 ($e$ = 1, 2, 2.5, 3) based on the aspect ratio of the osteons in cortical bone (1.02 < $e$ < 2.62), where $e$ = 1 represent a circle. To understand the presence of a potential relation between the spatial arrangement of the tubes and crack path, in the circular and elliptical designs, the degree of disorder was characterized and quantified by using two scalar quantities ($T$, $Q$) and radial distribution functions using a statistical mechanics approach in comparison to the theoretical perfectly random (ideal gas) and a perfectly ordered (HCP crystal) in two dimensions (**Figure S1**, Supporting information). It was found that although the circular and elliptical designs are periodic, they are not perfectly ordered, and thus have translational ($T$) and orientational ($Q$) order parameters below 1 (1 represents a perfect crystal) (**Figure S2**, Supporting information).

We characterize the fracture response and MOR of the tubular architected cement-based materials using single-edge notched bending (SENB) and three-point bending test, respectively, for various designs (i.e., circular and elliptical) and porosities ($\varphi$ = 20%, 30%, 40%, and 50%) in comparison with the reference solid counterparts fabricated with the same material composition (**Figure 3, Figure S5, Figure S6,** Supporting Information). More specifically, Figure 3 represents specific load vs displacement plots obtained from the SENB test, specific crack initiation and crack propagation fracture toughness, resistance curve, and specific MOR for the case of circular ($\varphi$ = 40% & 50%) as well as elliptical ($\varphi$ = 40%, $e$ = 2.5) designs. Circular, 40% and Elliptical, 40% ($e$ = 2.5) are chosen for conciseness and detailed discussion as they exhibit the most promising fracture response among the respective circular and elliptical designs of the tubular architected materials.

In comparison to the conventional single brittle failure in the solid counterparts (shown in grey), the specific-load vs. displacement response of tubular architected material obtained from the SENB test (Figure 3a), highlights a unique globally non-brittle and non-abrupt fracture behavior (shown in red and blue). The global non-brittle behavior is characterized by a postponed abrupt failure after the initial peak load provided by the circular and elliptical tubular architecture. Unlike the catastrophic failure at the first peak load in solid brittle materials, the tubular architected designs postpone the abrupt failure and exhibit multiple load drop and increase steps hence allowing for an overall hardening and softening behaviors after the first peak.

Analysis of the fracture profiles for Circular, 40% (Figure 3b, **Movie S1**) and Elliptical, 40% (Figure 3c, **Movie S2**), obtained through Digital Image Correlation (DIC), reveals more insight into the relationship between the multi-step attributes of the load vs. displacement curve (Figure 3a) and the crack path (Figure 3b,c). The first load drop in the response of Circular, 40% and Elliptical, 40% (Figure 3a) is shown in blue and red, respectively, and corresponds to crack initiation from the sharp notch to the tube in the bottom-most row in these two architectures (Figure 3b, i-ii & Figure 3c, i-ii). This is followed by a common increase in the load in both the elliptical and circular cases (Figure 3a) and indicates the reinitiation of the crack from the first tube ahead of the initial notch and propagation toward the subsequent row of tubes, respectively (Figure 3b, ii-iii & Figure 3c, ii-iii). The subsequent load drops in the load-displacement response (Figure 3) continue to correspond to the additional interaction of the crack with the tubes ahead of the crack tip in both types of designs (Figure 3b, iii-v & Figure 3c, iii-v). This unique behavior underscores the remarkable ability to pin the crack to the tube that acts to stabilize the overall crack propagation by promoting crack-tube interaction. This phenomenon gives rise to the overall non-brittle non-abrupt response in the circular and elliptical tubular architected cement-based materials compared to the cast counterparts.

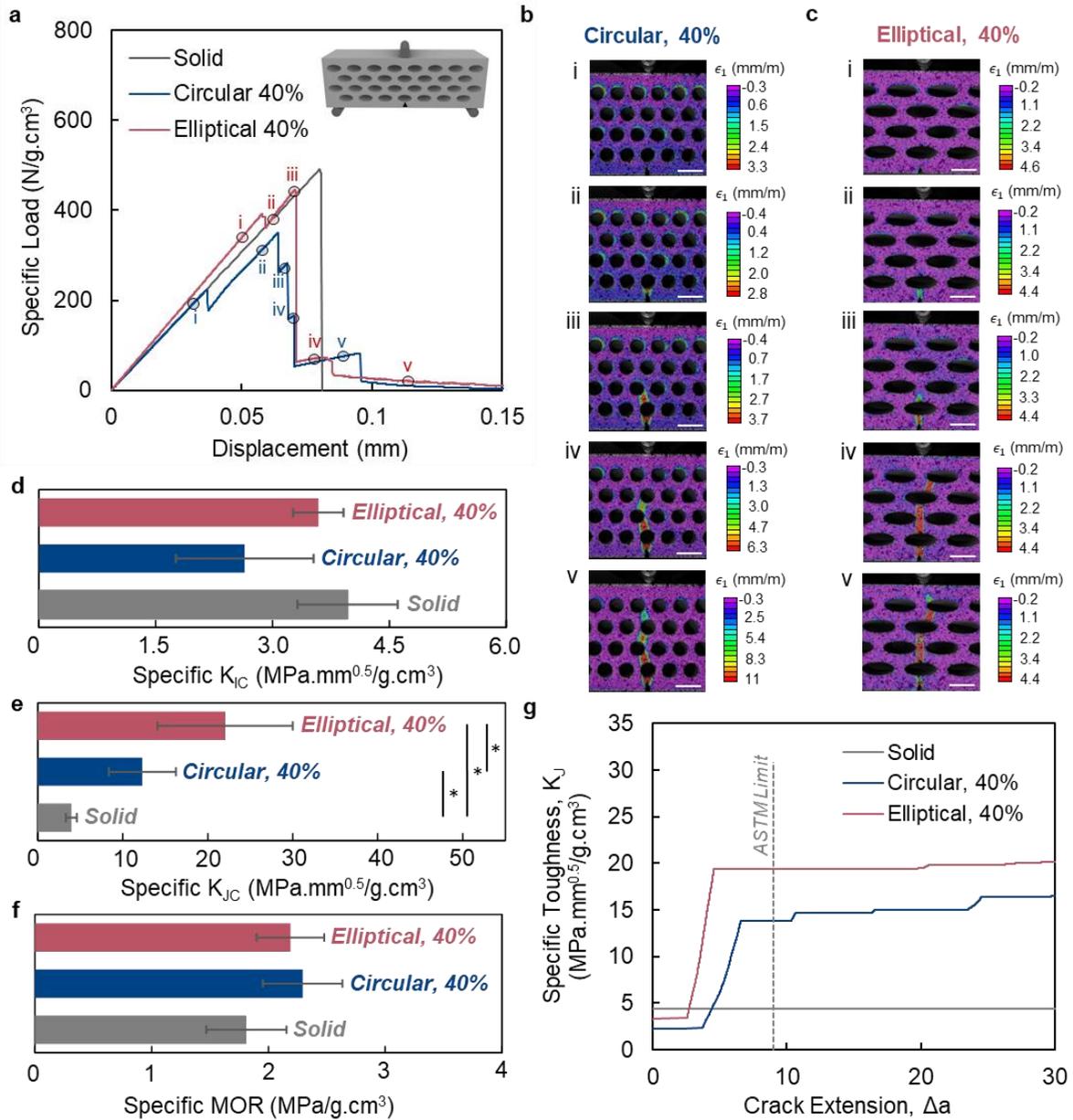

**Figure 3. Mechanical response of circular and elliptical tubular architected cement-based materials compared to the solid counterparts**. a) Specific load vs. displacement curves from the SENB test with the corresponding digital image correlation (DIC) analysis based on maximum principal strain ($\epsilon_1$) of b) Circular, 40% and c) Elliptical, 40% designs. d, e) Comparison of the specific crack initiation fracture toughness ($K_{IC}$) and crack propagation fracture toughness ($K_{JC}$) of tubular and solid materials. f) Specific modulus of rupture (MOR) of architected and monolithic solid materials obtained from the three-point bending test. g) Specific

toughness (calculated using J-integral) vs. crack extension (Δa) demonstrating rising R-curves for architected tubular materials compared with the constant fracture toughness of monolithic solid counterpart. Data is shown as mean ± SD. '∗' depicts $p < 0.05$ which indicates the statistically significant difference between the samples (at the ends of the solid line below a '∗'). *p*-value is obtained from the F-test and T-test.

The fracture response of tubular architected materials and the solid counterparts is presented in terms of specific crack initiation fracture toughness ($K_{IC}$) and specific crack propagation fracture toughness ($K_{JC}$), as illustrated in Figure 3d and Figure 3e, respectively. The Circular, 40% and Elliptical, 40% display the specific $K_{IC}$ of $2.64 \pm 0.88$ MPa.mm$^{0.5}$/g.cm$^3$ and $3.59 \pm 0.32$ MPa.mm$^{0.5}$/g.cm$^3$ which is statistically similar to the solid cement-based material, $3.96 \pm 0.64$ MPa.mm$^{0.5}$/g.cm$^3$ (in the confidence level of 90%). A similar $K_{IC}$ normalized with respect to density indicates that the presence of tubular architecture does not lead to any significant loss of resistance to cracking.

In addition to specific $K_{IC}$, the specific crack propagation fracture toughness ($K_{JC}$) can be characterized to capture the resistance to cracking in cases where the material does not abruptly fail in a brittle manner upon crack initiation. The Circular, 40% and Elliptical, 40% exhibit the specific $K_{JC}$ of $12.30 \pm 3.96$ MPa.mm$^{0.5}$/g.cm$^3$ and $22.04 \pm 7.96$ MPa.mm$^{0.5}$/g.cm$^3$ corresponding to the ultimate resistance that takes place at the point of complete failure of the notched specimens. These values of $K_{JC}$ for circular and elliptical architected materials are significantly higher than those of solid counterparts by 3.1 and 5.6 folds, respectively. Furthermore, the ASTM limit of $K_{JC}$ of Elliptical, 40% is $20.48 \pm 8.71$ MPa.mm$^{0.5}$/g.cm$^3$ is significantly higher than that of solid material. Circular, 40% demonstrates the ASTM limit of $K_{JC}$ as $8.92 \pm 5.28$ MPa.mm$^{0.5}$/g.cm$^3$, which is statistically similar to its solid counterpart.

In addition, the specific MOR of the Circular, 40%, and Elliptical, 40% are $2.22 \pm 0.33$ MPa/g.cm$^3$ and $2.12 \pm 0.28$ MPa/g.cm$^3$, respectively, which are statistically similar to that of the solid material ($1.76 \pm 0.33$ MPa/g.cm$^3$) as shown in Figure 3f. The findings show that engineering the tubular architecture to enhance the specific fracture toughness of the brittle cement-based material can be achieved while maintaining its specific MOR (i.e., the presence of the tubes has not led to additional loss of strength). The results assert the ability to overcome the often mutually exclusive nature of these two properties in brittle materials by developing a stable crack propagation through leveraging tubular architecture.

The R-curve plot captures the specific fracture toughness as a function of crack growth ($\Delta a$) in tubular architected material in comparison to the solid material (Figure 3g). Therefore, the elegance of the R-curve is in demonstrating the entire fracture process and the crack initiation, stepwise cracking through interaction with the circular and elliptical tubes, and overall crack propagation path until ultimate failure. In comparison to the brittle solid cement-based materials that typically demonstrate a brittle failure upon crack initiation (only a constant $K_{IC}$) as illustrated with a constant value on the Y axis with a gray line in Figure 3g, architected tubular material of the same composition represent a drastically dissimilar resistance curve. The R-curve demonstrates that the specific fracture toughness of tubular architected material increases in a stepwise manner where an iterative series of rapid crack extensions ($\Delta a$) on the x-axis are followed by the increase in the specific fracture toughness on the y-axis. These stepwise crack extensions from the fracture analysis used for the R-curve (based on the load drop) are compared with the experimentally observed crack extensions (from the DIC) and represent a great agreement (**Figure S7**, Supporting Information). In other words, this stepwise rise in the R-curve is directly correlated with the hardening and softening behavior observed as the load increase and load drop in the load-displacement plot in Figure 3a at the events of crack initiation from the notch and the subsequent reinitiation from the tubes (Figure S7, Supporting Information). In contrast to the flat resistance curve in cast brittle cement-based tubular architected counterparts of the same constituents provide unique rising R-curves and higher fracture toughness upon initiation.

3. **Discussion**

The interaction of the crack with the tube and the subsequent stepwise cracking is the mechanism for toughening tubular architected material of brittle constituents. Stepwise cracking involves a mechanism in which the tube first arrests the crack, followed by the crack reinitiation from the tube. Corroboration of crack-tube interaction using DIC and the fracture toughness calculations from the load-displacement provides evidence for such a characteristic of stepwise cracking (Figure 3a,b,c and Supporting information Figure S7). For instance, the five and four steps in crack extension based on the R-curve very well match the five and four steps in the observed crack for the Circular 40% and Elliptical 40% tubular material, respectively (Supporting information Figure S7). This continued and stepped crack propagation ultimately provides the rise in the R-curve, additional stepwise crack extension (Figure S7a,c, Supporting Information), and higher fracture toughness in tubular architected cement-based materials.

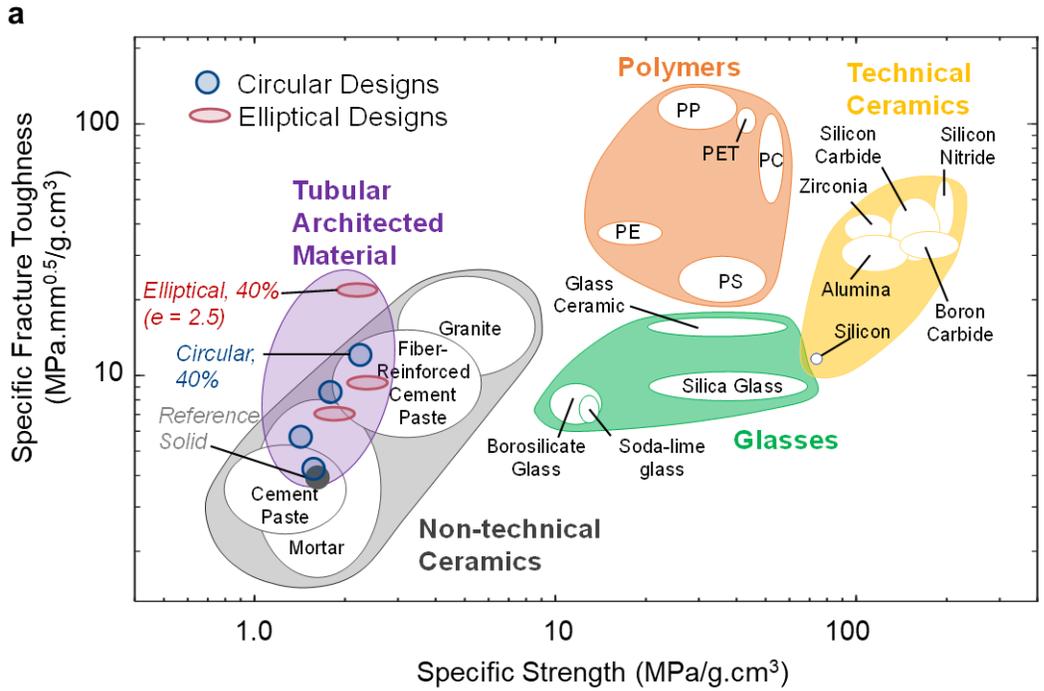

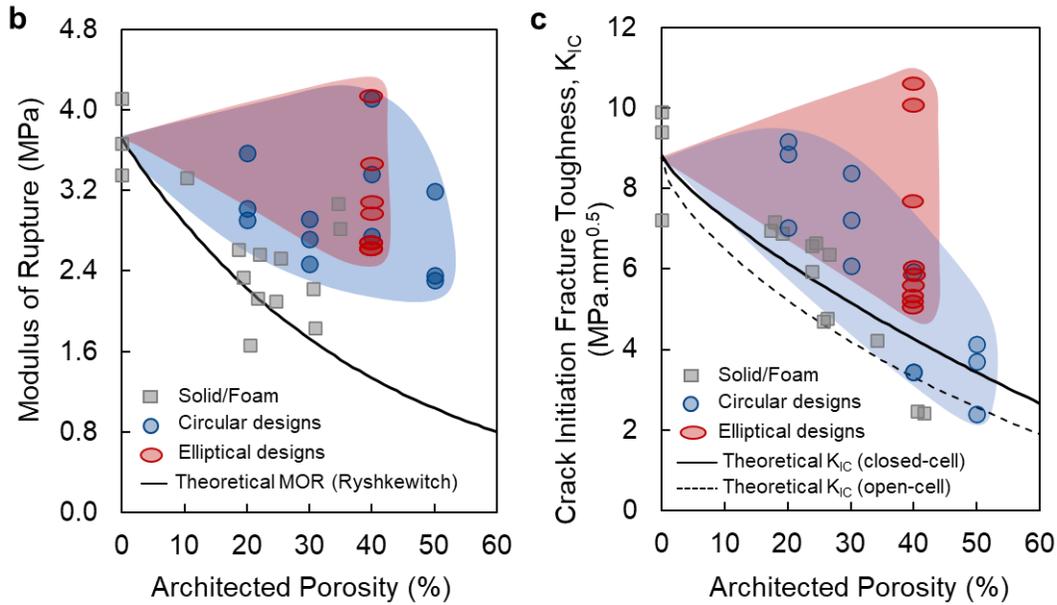

**Figure 4. Ashby plot and modulus of rupture of architected material.** a) Ashby plot displaying specific fracture toughness vs. specific strength, featuring tubular architected cement-based materials in comparison to our solid cement paste as well as the survey of the literature from cement paste, [35,37,65–70] fiber-reinforced cement paste, [71–75] and mortar [35,37,67,73] alongside various engineering materials, where the most promising architected materials exhibit higher specific fracture toughness than solid cast, and even mortar, and fiber-reinforced cement paste counterparts, all while maintaining their specific strength. b, c) MOR and $K_{IC}$ of tubular architected cement-based materials with varying porosity ($\varphi$ = 20%, 30%, 40%, & 50%) and aspect ratio ($e$ = 1, 2, 2.5, & 3) compared to reference solid cement-based material and theoretical MOR and $K_{IC}$ (for open-cell and closed-cell material) obtained using the Ryshkewitch relation[76] and geometric model, respectively. [77]

The Ashby plot (**Figure 4**a) demonstrates the comparison of the specific fracture toughness vs. specific strength of all investigated designs of tubular architected against the solid cement paste and fiber-reinforced cement paste materials and mortar, as well as other engineered materials such as glass, polymers, and technical ceramics.[78] The fracture toughness and flexural strength of solid reference cast closely lie within the (upper limit) range of values obtained from the literature. The circular designs with smaller porosity ($\varphi$ = 40% and 50%) exhibit slightly higher specific fracture toughness compared to solid cement-based material. However, the circular and elliptical designs with higher porosity ($\varphi$ = 40% and 50%) demonstrate 3 and 2 folds higher fracture toughness, respectively, than the solid cement-based representing the conventional properties of cement paste, placing them in the same toughness range as the fiber-reinforced cement paste. Specifically, the most effective design of tubular architecture, the Elliptical, 40% shows a specific fracture toughness of 22.04 $\pm$ 7.96 MPa.mm$^{0.5}$/g.cm$^3$, which exceeds the specific toughness of the cementitious materials found in the literature for conventional cement paste (3.89 $\pm$ 1.5 MPa.mm$^{0.5}$/g.cm$^3$),[35,37,65–70] mortar (4.94 $\pm$ 3.5 MPa.mm$^{0.5}$/g.cm$^3$),[35,37,67,73] and fiber-reinforced cement paste (10.12 $\pm$ 4.2 MPa.mm$^{0.5}$/g.cm$^3$).[71–75]

On the other hand, compared to the theoretical MOR obtained using the Ryshkewitch relation[76] and reference solid foam cast material, the circular and elliptical tubular architected materials display a relatively higher MOR (Figure 4b, Figure S5e, Figure S6e, Supporting Information). Similarly, we have compared the crack initiation fracture toughness ($K_{IC}$) of tubular architected materials alongside that of foam cement and the theoretical $K_{IC}$ values for open-cell and closed-cell materials (where the cell represents the pore phase in a geometrically conceptualized model[77]), as shown in Figure 4c. Tubular architected materials with circular designs generally outperform the solid foam in terms of $K_{IC}$ at the examined range of (architected and entrained) porosity. Additionally, the circular architected materials exhibit higher $K_{IC}$ at lower porosities (10 – 30%) when compared to the theoretical $K_{IC}$ of closed cell counterparts[77] and at higher porosities (40 – 50%), converge towards the theoretical value for $K_{IC}$ for open-cell materials[77]. Moreover, the elliptical tubular architected materials examined here, surpass solid foam materials, and both closed-cell and open-cell theoretical $K_{IC}$ [77] at comparable porosities.

Investigation of a broader range of the porosity in the circular architected materials and comparison against cast counterpart demonstrates an additional insight about the optimality of the 'tubular porosity' of $\varphi = 40\%$ case compared to other cases. The outperformance of the $\varphi = 40\%$ case in terms of fracture toughness is enabled by the effectiveness of the stepwise cracking toughening mechanism in this case. By conducting a theoretical analysis of the stress intensity factor, $K_I$ at the notch, due to the presence of a single tube ahead of the notch (**Figure S9-S11**, Supporting Information), the underlying phenomena about the relation between the stress field and stress intensity factor can be elucidated. It follows that as the tubular porosity, $\varphi$, increases by increasing the radius of the tube, the stress intensity factor at the notch due to the presence of the hollow tube increases leading to a lower load to initiate a crack (Figure S9c, d, Supporting Information). The increase in stress intensity factor results in the reduction in the stress and corresponding value of the peak load required for initiation of the crack from the notch tip as the tubular porosity increases ahead of the tip (Figure S9e, f, Supporting Information). Thus, at the largest (manufacturable) tube size $\varphi = 50\%$, the lowest fracture toughness is expected, given the relationship between load and fracture toughness. At the smallest tube size at $\varphi = 20\%$, the highest load is expected. However, the other key to enhanced fracture toughness is the ability to promote the crack-tube interaction, which gives rise to the cracking and rising R-curves.

These two mechanisms compete, as evident from the load-displacement plot. As the tubular porosity increases, although the fracture toughness should decrease, the larger tubes enable the crack to interact with the tube and provide crack arrest, which provides a pathway for more stable crack propagation and a rising R-curve. Although at large tube sizes $\varphi = 50\%$, the stepwise cracking remains and provides stepwise crack extension as seen in the R-curves and load vs CMOD plot (Figure 3g, Figure S5b, Figure S8a, Supporting Information), its effect diminishes as the critical load to initiate the crack significantly decreases with increasing tube size. On the other hand, at the smallest tube size $\varphi = 20\%$, the spare distribution of the voids due to their small size does not promote crack-tube interaction, though it may provide higher peak load and fracture toughness at the initiation. This gives rise to an optimal tubular porosity of $\varphi = 40\%$ (Figure S5d, Supporting Information), which was extended to an elliptical design with varied aspect ratio.

In terms of aspect ratio, the best performance of $e = 2.5$ in terms of fracture toughness is attributed to the competing stepwise cracking toughening mechanism and $e$. For a single tube interacting with a crack, an increase in $e$ leads to a decrease in the $K_I$ (Figure S9d, Supporting Information), which in turn necessitates an increase in the peak load required to initiate the crack (Figure S9e, f, Supporting Information). This finding suggests that the case of $e = 3$ is expected to have the highest fracture toughness, whereas the case of $e = 2$ is expected to have the lowest fracture toughness. However, fracture toughness also depends on the growth in the R-curve, which is a result of the stepwise cracking toughening mechanism. Thus, as stepwise cracking competes with $e$ for the peak required for crack initiation, the lower aspect ratio ($e = 2$ and $e = 2.5$) promotes stepwise cracking, in contrast to $e = 3$ which does not exhibit stepwise cracking, as shown by R-curve and load vs CMOD plots (Figure 3g, Figure S6b, Figure S8b, Supporting Information). Considering the peak load and the stepwise cracking toughening mechanism, $e = 2.5$ demonstrates the highest fracture toughness compared to all tubular designs (Figure 3e, Figure S6d, Supporting Information).

Here, we engineer the toughening mechanism of stepwise cracking, which includes crack arrest and crack reinitiation, in cement-based tubular architected material with an inherently brittle constituent inspired by the crack-cement line interaction in cortical bone. Engineering tubes as defects ahead of the crack in cement-based materials can help generate unique toughening mechanisms inspired by defect-crack interaction in cortical bone. It should be noted that the stepwise toughening mechanisms emerge when the volume fraction of tubular voids reaches 40 – 50% (of both circular and elliptical cases), a range that aligns closely with the 45 – 60% volume fraction of osteons in cortical bone. [56] Additionally, the most promising case of elliptical design with 40% porosity and aspect ratio $e$ = 2.5, is also within the matching range of the osteons' volume fraction and falls in the aspect ratio range of osteons in cortical bone ($e$ = 1.02 to 2.62).[57]

Further studies can investigate the arrangement of the tubes ahead of the crack. Using a statistical mechanics approach, further understanding of the seemingly periodic vs. non-periodic defects such as tubular porosity can be achieved by translating periodicity to 'order parameters'. This approach can help formalize a quantifiable metric into the design of materials arrangement (solid or pore) and assist in performing inverse design of tubular porosity. From a statistical mechanics perspective, two limits of random (ideal gas) and perfectly ordered (FCC crystal) can be constructed with a '*disorder*' spectrum in between.[79,80] From a design of architected materials perspective, a wide range of disorder arrangements (of tubular porosity or any other architecture and phase), bound by the random and ordered limits, can be conceptualized and mechanically probed in the context of fracture mechanics. In this work, a rigorous approach to characterize the arrangement of architected material is proposed for the first time with the preposition to move away from inadequate periodicity concepts and towards the mathematically conceivable representation of the materials using disorder (i.e., degree of disorder) in the field of architected materials. As such, we propose to capture 'disorder' using radial distribution function (**Figure S1, Figure S2, Figure S4,** Supporting Information) and more quantitatively, by related translational order parameter, $T$ (See Eq. S12), and orientational parameters, $Q$ (See Eq. S13) in this endeavor.

The random system is the lower bound for $T$ and $Q$, and the perfect FCC crystal (in 3D) or hexagonal closed packing (HCP) crystal (in 2D) is the upper bound for $T$ and $Q$, based on statistical mechanics.[79,80] In this study, given the two-dimensional nature of the tubular design, we suffice to investigate the order parameters for the proposed materials in 2D systems, leading to a range from an HCP lattice to random distributions.

Thus, within the 2D systems, we examine the translational order parameter ($T$) and the orientational order parameter ($Q = Q_6/Q_6^{HCP}$), for these limits leading to range from 1 to 0 for $T$ and from 1 (HCP) to 0.42 for $Q$ (equivalent to 0.74 to 0.31 for $Q_6$), respectively. This variation in $T$ and $Q$ parameters corresponds to a range from 'perfectly ordered' to 'highly disordered' systems (where $Q$ approaches 0 in 3D but not in 2D) and is illustrated in **Figure 5**a and 5b, respectively. This range is broad and encapsulates the degree of disorder as it exists in the arrangement of both tubular cases probed in this study as well as the natural materials. The tubular architecture in nature exhibits a high degree of disorder, with tubules in Dentin showing *T* and *Q* parameter values as 0.22 and 0.51, respectively, whereas osteons in cortical bone display *T* and *Q* parameter values as 0.02 and 0.47, respectively.

The rigorous framework for capturing disorder using radial distribution function, $g_2(r)$ (Figure S4) and *T* and *Q* order parameters, for a range of broad distributions from perfectly ordered to highly disordered (approaching random in 2D), were developed by considering a normalized step size (using metropolis algorithm – **Figure S3**a-d). This leads to obtaining what we refer to here as '*disorder curves*' representing the *T* and *Q* as a function of step sizes $\delta$, normalized by the first nearest neighbor distance for the HCP ($d_{HCP}$) as shown in Figure 5a, and 5b. It is worth noting that the order *T* can be quite sensitive to the statistical mechanics parameters used to calculate it (i.e., number of particles N, shell thickness $\Delta$, and the number of shells Ns). Thus, a thorough sensitivity analysis is conducted to allow for convergence of the T value to a plateau for each of these three parameters (See Figure S3e-i).

Additionally, we use the void centroid to represent the arrangement of both natural and engineered materials to calculate the corresponding *T* and *Q* parameters values and impose them on the 'disorder curves' (Figure 5a and b). The osteons in cortical bone and tubules in dentin exhibit the same degree of disorder as liquid (highly disordered) and glass ('somewhat disordered'), respectively, for both *T* (0.22 & 0.02) and *Q* (0.51 & 0.47) parameters.

The circular and elliptical tubular arrangements proposed in this work represent a translational order parameter, T of 0.21 and 0.30 (close to those dentin), and orientational parameters, Q of 0.97 and 0.60, respectively. This reveals that although the proposed circular and elliptical architected material appears periodic, the T parameter meaningfully represents a value corresponding to a 'somewhat disordered' range of materials (Figure 5a).

In contrast, the proposed tubular architected material represents drastically different *Q* parameters between the circular and elliptical designs, indicating the utility of Q in the slight nuances of the design of the arrangement (Figure 5b). In addition, the *Q* parameter for the proposed circular architected materials represents an intuitively meaningful value of Q = 0.97, very close to a perfectly ordered HCP system (Q = 1), as depicted in Figure 5b. The *Q* parameter for the elliptical architected materials, on the other hand, represents a value of Q = 0.60, in the range between 'somewhat disordered' and 'highly disordered', as illustrated in Figure 5b.

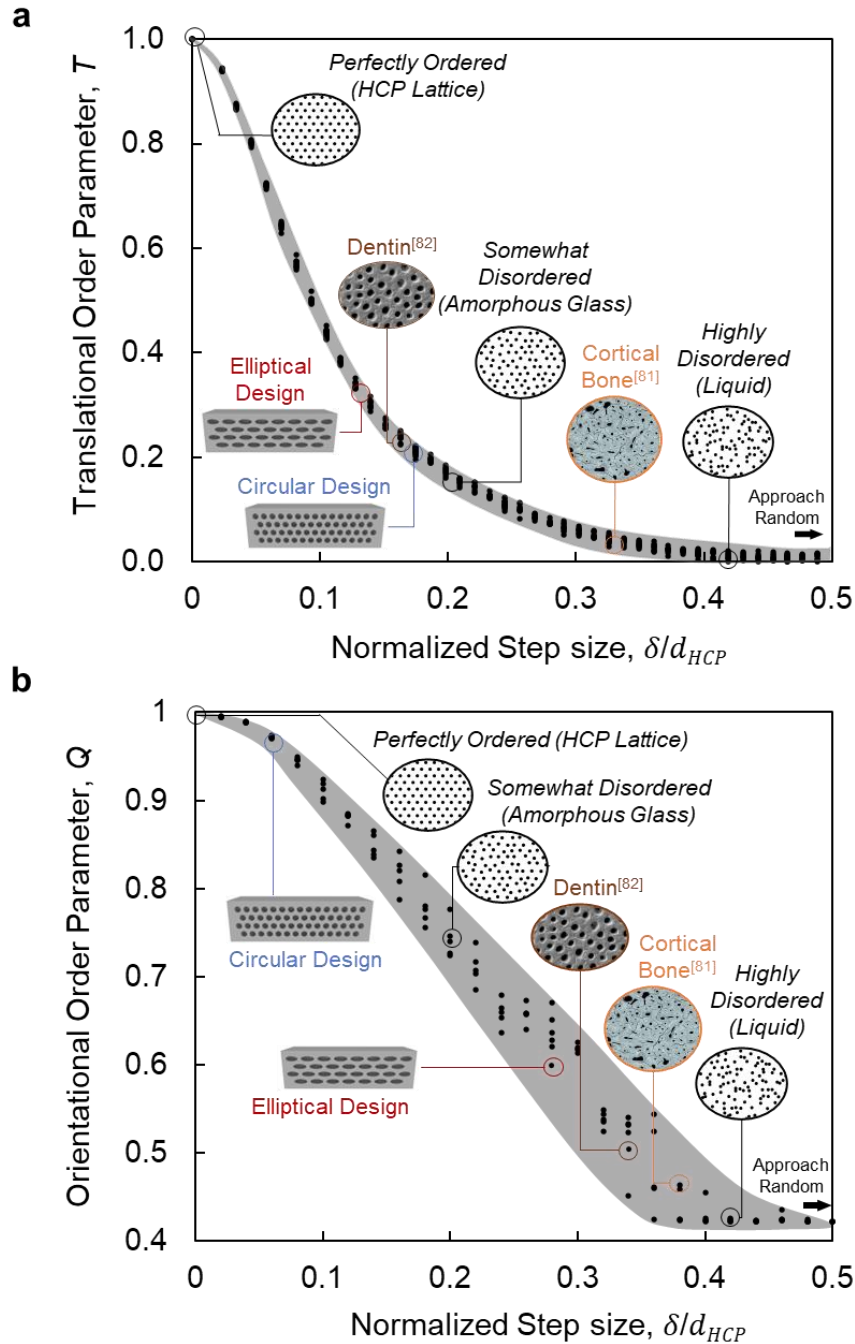

**Figure 5. Disorder Curves Representing Order Parameters (T, Q) vs. Normalize Step Size based on a Statistical Mechanics Approach, imposed on them the Order Parameter Values of Natural and Engineered Tubular Architected Materials.** a) Translational order parameter

($T$) and b) orientational order parameter ($Q$) with varying step size ($\delta/d_{HCP}$) for 2D – distributions, ranging from perfectly ordered, to disordered ('somewhat' and 'highly' corresponding to amorphous glass and liquid), ultimately approaching random. The order parameters were independently computed based on the distribution of centroids of tubes in tubular architected material (Circular and Elliptical arrangement) along with the distribution of the centroids of osteons in cortical bone adapted from [81] and tubules adapted from [82] in Dentin are presented on $T$ vs. $\delta/d_{HCP}$ and $Q$ vs. $\delta/d_{HCP}$ disorder curves. The step size ($\delta/d_{HCP}$) controls the degree of disorder in the system with $\delta = 0$ indicating the perfect lattice and $\delta = 0.5d_{HCP}$ as the highly disordered system). The variation in the $T$ and $Q$ values for the fixed value of step size ($\delta/d_{HCP}$) is statistical noise, which is generated due to the stochastic nature of the Metropolis algorithm used to generate the disordered pattern. More specifically, the random sampling of displacement of particle within the maximum allowable step size ($\delta$) in the Metropolis algorithm (Figure S3, Supporting Information) generates this statistical noise. The gray strips in $T$ vs. $\delta/d_{HCP}$ and $Q$ vs. $\delta/d_{HCP}$ disorder curves encompass the statistical noise generated from 10 repetitions.

These findings highlight that neither '*periodicity*' is equivalent to '*ordered*' in a system, nor is '*non-periodicity*' equivalent to '*disorder*', highlighting a mechanically robust framework can provide better insight into the nature of a given design and the design-performance relationship. Rather the arrangement of phases in architected materials that can take any distribution other than perfect HCP lattice exhibit a certain degree of disorder considering either $T$ or $Q$ order parameters. Thus, the proposed set of order parameters can be powerful in tagging a given design with a quantifiable arrangement metric and probing for various mechanical and functional characteristics. The proposed algorithm is powerful in that it can be applied to any architected arrangement of materials beyond those imposed in Figure 5 here (natural and engineered tubular materials).

The findings highlight the effectiveness of $Q$ and $T$ in capturing the degree of disorder (e.g., heterogeneity) quantitatively in tubular and, more broadly, in the field of architected materials (Figure S1, S2, S3, S4 Supporting Information). More importantly, the proper application of statistical mechanics and 'degree of disorder' can help suggest a better measure than the assessment via 'periodicity'.[83–85] The binary periodic vs. non-period classical categorization of architected materials[83–85] falls short of capturing the spectrum in either period or non-period material and the role such spectrum plays in the material design, the mechanical properties or the interacting characteristics of the material (crack, wave, etc.) (Figure S1, S2, S3, S4 Supporting Information).

## 4. Conclusion

In conclusion, conventional cement-based materials exhibit brittle failure under fracture due to limited toughening mechanisms such as uncracked ligament bridging[86] and microcracking.[87] As such microstructure of concrete suffers from limited energy absorption capacity and low resistance to cracking (low fracture toughness). This has limited the reliance on tensile and flexural properties of cement-based materials, has imposed design constraints (reliance on and designing for compressive properties), and has imposed limits on the design landscape. Here, we propose engineering intentional tubular defects into cement-based materials to tailor the properties of monolithic counterparts to enhance the specific fracture toughness. The underlying competition between tube size and shape enables the engineering of stepwise cracking, which in turn significantly enhances fracture toughness through rising R-curves.

The findings of this study can, in principle, be extended to quasi-brittle mortar and concrete, contributing to the formulation of a strategy for enhancing the fracture toughness of concrete materials through deliberate design and harnessing defects, enabling the creation of crack-resistant characteristics in brittle and quasi-brittle cement-based materials. Deployment of these types of designs across size scales can be achieved, for instance, through leveraging the growing additive manufacturing techniques with cement paste at the desktop level to mortar and concrete at the robotic fabrication level.[88–90] The use of statistical mechanics can help quantify the spectrum of disorder in the material, thus enabling a higher design level, as a means to engineer certain interaction characteristics with the material.

A statistical mechanics approach to quantifying disorder in architected materials can inform a novel approach to the design of not only the material but also the degree of disorder it contains. It can promote the qualitative assessment of the design-performance relationship as it can elegantly help capture the orientation and translational aspects of geometry. Thus, this approach can be insightful for understanding the mechanics-geometry relationships across various classes of architected materials (from brittle to ductile) and meta-materials that interact with a field (stress, wave, damage, etc.), in an intuitive or exhaustive fashion. The utility of this approach and proposed algorithm can be further unleashed by broadly or specifically applying it to understanding geometry in the first place or using it to design and achieve certain disorders to reverse engineer the mechanics of interest. Using a quantifiable and mathematically conceivable representation of geometry we can, for instance, more specifically answer questions with regard to optimality for strong, tough, or damage-resilient designs in extreme or ultimate states in a more refined fashion in the context of the arrangement of material(s) or phase(s). These questions regarding the mechanics of material have generally been asked in the context of periodicity and non-periodicity. Nevertheless, they may be revisited from an order/disorder design perspective to discover new (bio-inspired) design principles and insights.

## 5. Experimental Section

*Preparation of cement paste*: Cement paste is composed of commercially available Type I cement acquired from Buzzi Unicem (Stockertown, PA, USA), deionized water, high range water reducing admixture (HRWRA, MasterGlenium 7700), and viscosity modifying admixtures (VMA, MasterMatrix 362). The cement paste is prepared by mixing 250 g cement, 71.5 g water, 1.84 g HRWRA, and 2.38 g VMA using the Twister Evolution Venturi vacuum mixer (Renfert®). The mixing is performed using a two-step mixing procedure.[88,89] The first step involves pre-mixing for 25 s, followed by mixing at 400 rpm for 90 s at 70% vacuum. The second step of mixing was conducted at 400 rpm for 90 s at 100% vacuum.

*Hybrid Manufacturing*: The tubular specimens are fabricated employing hybrid 3D-printing and casting techniques (Figure 2). The first step involves 3D-printing of positive mold using Polyvinyl alcohol (PVA) with the Ultimaker S5 as shown in Figure 2a. Subsequently, two-component urethane rubber (VytaFlex™ 40) was mixed in 1:1 by weight. The mix is then poured into the PVA mold to form the negative mold of the tubular specimen. Immediately following the pouring, the PVA mold filled with liquid urethane rubber is placed into a vacuum chamber for 10 minutes to extract the air bubbles from the liquid urethane rubber. The urethane rubber is then left to harden for at least 14 h at 35 ± 5% relative humidity and 25 ± 3 °C. The urethane rubber mold is then obtained by keeping the molds in the ultrasonic bath at 90° C for 12 hours to dissolve the PVA (Figure 2b). The fresh cement paste is then poured into the urethane rubber mold to cast tubular architected materials (Figure 2c). The specimens are fabricated in a laboratory environment at 21 ± 3 °C and relative humidity of 45 ± 5%. Then, they are transferred to a curing chamber maintained at a relative humidity of 97 ± 2% using saturated solutions of potassium sulfate,[91,92] where they undergo a 7-day curing process.

*Mechanical Properties Characterization*: The modulus of rupture (MOR) is characterized using the three-point bending test (3PB) according to ASTM C293M-16.[93] The fracture toughness is determined using the single-edge notch bend (SENB) according to the following ASTM E1820-20b.[94] The unnotched and notched prismatic specimens of 40 × 40 × 130 mm dimensions are prepared for 3PB and SENB testing, respectively. In SENB specimens, a 2 mm deep notch was first introduced using a 2 mm thick diamond saw followed by an additional 2 mm depth using the razor of 0.3 mm thickness. The loading rates of 0.1 mm/min and 0.05 mm/min are employed for 3PB and SENB testing, respectively. Both mechanical tests are performed using the electromechanical frame (MTS Criterion C45.305) with a load cell of 20 kN capacity. At least three repetitions of each type of specimen are performed and the average value is reported. The details of the calculations involved for each test are discussed in Supporting Information.

*Statistical Analysis.* Data is presented as mean ± SD. A minimum of 3 repetitions were performed for each test. The one-tailed F-test and T-test with a confidence interval of 95%, were employed as a statistical method to study the significant difference between variance and mean of the data, respectively. $p < 0.05$ indicates the statistically significant difference between the samples. $\alpha$ value of 0.05 was used. Data Analysis toolbox in Microsoft Excel was used to perform statistical analysis.

**Supporting Information**

Supporting Information is available from the Wiley Online Library.

**Conflict of Interest**

The authors declare that they have no competing interests.

**Acknowledgments**

We would also like to thank the Buzzi Unicem for providing the cement and BASF for chemical providing chemical admixture. This work was partly supported by the National Science

Foundation CAREER Award (2238992) and the partial support from National Science Foundation CMMI Division Grant (ECI, 2129566).

*Authors Contributions*: SG and RM conceived the project, SG conducted the experiments and analysis of data, SG and RM interpreted the data, SG and RM wrote and edited the original manuscript, and RM acquired funding.

Supporting Information

**Tough Cortical Bone-Inspired Tubular Architected Cement-based Materials**

*Shashank Gupta, Reza Moini\**

Department of Civil and Environmental Engineering, Princeton University, Princeton, NJ, USA

Email: Reza.Moini@princeton.edu

**Characterization of Fracture Toughness and Modulus of Rupture**

*Fracture Toughness*: The fracture toughness is determined using the single-edge notch bend (SENB) according to the ASTM E1820-20b.[1–10] The crack-initiation fracture toughness, $K_{Ic}$, is defined by Eq. S1.

$$K_{Ic} = \frac{P_c L}{W D^{1.5}} * f\left(\frac{a_o}{D}\right) \qquad (Eq. S1)$$

where, $P_c$ is the load corresponding to the initiation of the crack. $L$, $W$, $D$, and $a_o$ denote the span length, width, depth, and notch length of the prismatic beam, respectively. The $L$, $W$, and $D$ of the beams are 120 mm, 40 mm, and 40 mm, respectively. The notch length-to-depth ratio ($\frac{a_o}{D}$) of 0.1 is employed.[11] The function $f\left(\frac{a_o}{D}\right)$ considers the geometrical parameter (shape function) of the SENB specimen and the notch and is defined by Eq. S2. The $L/D$ ratio of 3.0 was used here which was considered for adjusting the value of shape function.[5,7,8,12,13]

$$f\left(\frac{a_o}{D}\right) = \frac{3\left(\frac{a_o}{D}\right)^{\frac{1}{2}}\left(1.99 - \left(\frac{a_o}{D}\right)\left(1 - \frac{a_o}{D}\right)\left[2.15 - 3.93\left(\frac{a_o}{D}\right) + 2.7\left(\frac{a_o}{D}\right)^2\right]\right)}{2\left(1 + \frac{2a_o}{D}\right)\left(1 - \frac{a_o}{D}\right)^{\frac{3}{2}}} \qquad (Eq. S2)$$

The *J*-integral determines the strain energy release rate for the crack in the material. The computation of the *J*-integral involves two components: the elastic component ($J_{el}$) and the plastic component ($J_{pl}$) The $J_{el}$ and $J_{pl}$ are defined by Eq. S3 & S4, respectively.

$$J_{el} = \frac{K_{Ic}^2}{E'} \qquad (Eq. S3)$$

where, $K_{Ic}$ is the crack initiation fracture toughness. $E'$ is the plane strain Young's Modulus of the hardened cement paste (30 GPa).[14–16]

$$J_{pl} = \frac{\eta_{pl} A_{pl}}{W(D - a_0)} \qquad (Eq. S4)$$

where $\eta_{pl}$ is the dimensionless function of geometry and is defined as 1.9 when load-line displacement is used for SENB specimen as per ASTM E1820-20b, $(D - a_0)$ denotes the remaining ligament length, and $A_{pl}$ is the area under the force-displacement curve beyond the load corresponding to the initiation of the crack ($P_c$) as shown by Eq. S5.

$$A_{pl} = \sum_{i=P_c}^{n} P_i \times \delta_i \quad (Eq. S5)$$

The fracture toughness, $K_J$, can be further defined as shown in Eq. S6.

$$K_J = \sqrt{(J_{el} + J_{pl})E'} \quad (Eq. S6)$$

The $K_J$ can be plotted for the incremental points of crack extension, $a_i$, which is defined by Eq. S7.

$$a_i = a_{i-1} + \frac{D - a_i}{2} \frac{C_i - C_{i-1}}{C_i} \quad (Eq. S7)$$

where, $C_i$ is the instantaneous compliance of the notched specimen and it can be defined by Eq. S8 as the function of the ratio of crosshead displacement ($u_i$) to applied load ($P_i$).

$$C_i = \frac{u_i}{P_i} \quad (Eq. S8)$$

The extension of the crack tip corresponds to an increase of $A_{pl}$ which allows plotting of the growth of fracture toughness against crack extension $a_n$ as a resistance curve (R-curve). The representation of fracture toughness in relation to the crack extension (R-curve) plot is valuable in providing an understanding of the evolution of the material's resistance to crack initiation and propagation.

According to ASTM E1820-20b [1], the maximum J-integral capacity is defined by Eq. S9.

$$J_{max} = min\left\{\frac{W\sigma_y}{10}, \frac{(D-a_0)\sigma_y}{10}\right\} \quad (Eq. S9)$$

On the other hand, the maximum crack extension for the specimen is defined by Eq. S10 according to ASTM E1820-20b.[1]

$$a_{max} = 0.25(D - a_o) \qquad (Eq. S10)$$

The $J$ value corresponding to the maximum crack extension ($a_{max}$) is lower than the $J_{max}$ (obtained from Eq. S9) for all the monolithic and tubular architected specimens in this study. The $J$ value corresponds to $a_{max}$ is used to determine the ASTM limit of fracture toughness of the material.

*Modulus of Rupture*: The modulus of rupture, MOR, is determined using the three-point bending test (3PB) of the unnotched prismatic specimen according to ASTM C293M-16.[17] The maximum peak load, $P_{max}$, is used to calculate the MOR as shown by Eq. S11.

$$MOR = \frac{3(P_{max})L}{2WD^2} \qquad (Eq. S11)$$

where, $L$, $W$, and $D$ denote the span length, width, and depth of the prismatic beam, respectively. The $L$, $W$, and $D$ of the beams are 120 mm, 40 mm, and 40 mm, respectively.

**Characterizing the *Degree of Disorder* in Architected Materials Using Statistical Mechanics**

Tubular architected material can be designed with various arrangements. In this study, the spatial distribution of the tubes is characterized and quantified using the radial distribution function and two-order parameters. The radial distribution function ($g_2(r)$) is the probability distribution of locating the point at a radial distance ($r$) from the center of a reference point similar to those applied in the molecular structure of gas and liquids.[18] Furthermore, the degree of disorder is quantified by two order parameters (translational and orientational) both of which are scalar quantities that represent the value of zero for a completely random distribution (e.g., particle distribution in an ideal gas) and 1 for a perfectly ordered distribution (e.g., crystal lattice).[18–21] Translational and orientational order parameters have been widely used to study the spatial distribution of molecules in liquid crystal and amorphous solids,[19,20] but have rarely been adopted to rigorously characterize a materials architecture beyond the qualitative periodic/non-periodic metrics.

The translational order parameter ($T$) measures the spatial ordering of particles relative to the perfect Hexagonal closed packing (HCP) lattice and the ideal gas, at the same particle density, and is defined by Eq. S12.[18]

$$T = \left| \frac{\sum_{i=1}^{N_s}(n_i - n_i^{ideal})}{\sum_{i=1}^{N_s}(n_i^{HCP} - n_i^{ideal})} \right| \quad (Eq. S12)$$

where, $n_i$ denotes the average occupation number for the shell of thickness $\Delta d_{HCP}$ which is centered at the distance from a reference sphere that equals the distance of the $i$th nearest neighbor for the HCP lattice of the same particle density; $d_{HCP}$ is the first nearest-neighbor distance for the HCP lattice; $N_s$ is the total number of shells; and $n_i^{HCP}$ and $n_i^{ideal}$ are the occupation numbers of the HCP lattice and ideal gas, respectively.

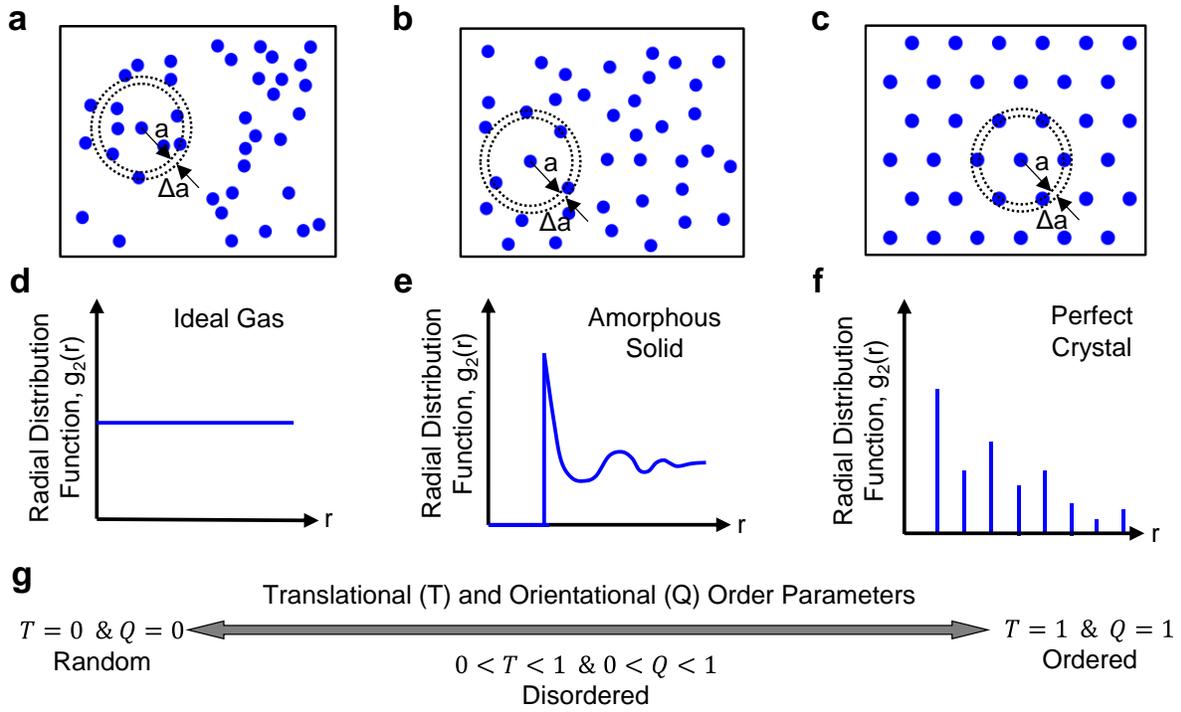

**Figure S1**. a-c) Random, disordered, and ordered distribution of particles, respectively, and the corresponding d-f) schematic of radial distribution functions, and g) the range of translational and orientational order parameters.

On the other hand, the orientational order parameter is defined by spherical harmonics ($Y_{lm}$), which are used in statistical mechanics to describe the angular distribution of particles. The orientational bond parameter, $Q_6$, (where, $l = 6$) is defined by Eq. S13.[20,21]

$$Q_6 = \left(\frac{4\pi}{13}\sum_{m=-6}^{6}|\bar{Y}_{6m}|^2\right)^{1/2} \quad (Eq.\,S13)$$

The orientational order parameter, $Q$, is defined as the orientational bond order parameter, $Q_6$, normalized by its value for perfect hexagonal close packing (HCP) lattice, $Q_6^{HCP}$. The orientational order parameter, $Q$, is expressed by Eq. S14.

$$Q = \frac{Q_6}{Q_6^{HCP}} \quad (Eq.\,S14)$$

The radial distribution function and the corresponding two-order parameters ($T$, $Q$) for random (ideal gas), disordered, and perfectly ordered (HCP) distribution are shown in **Figure S1**.

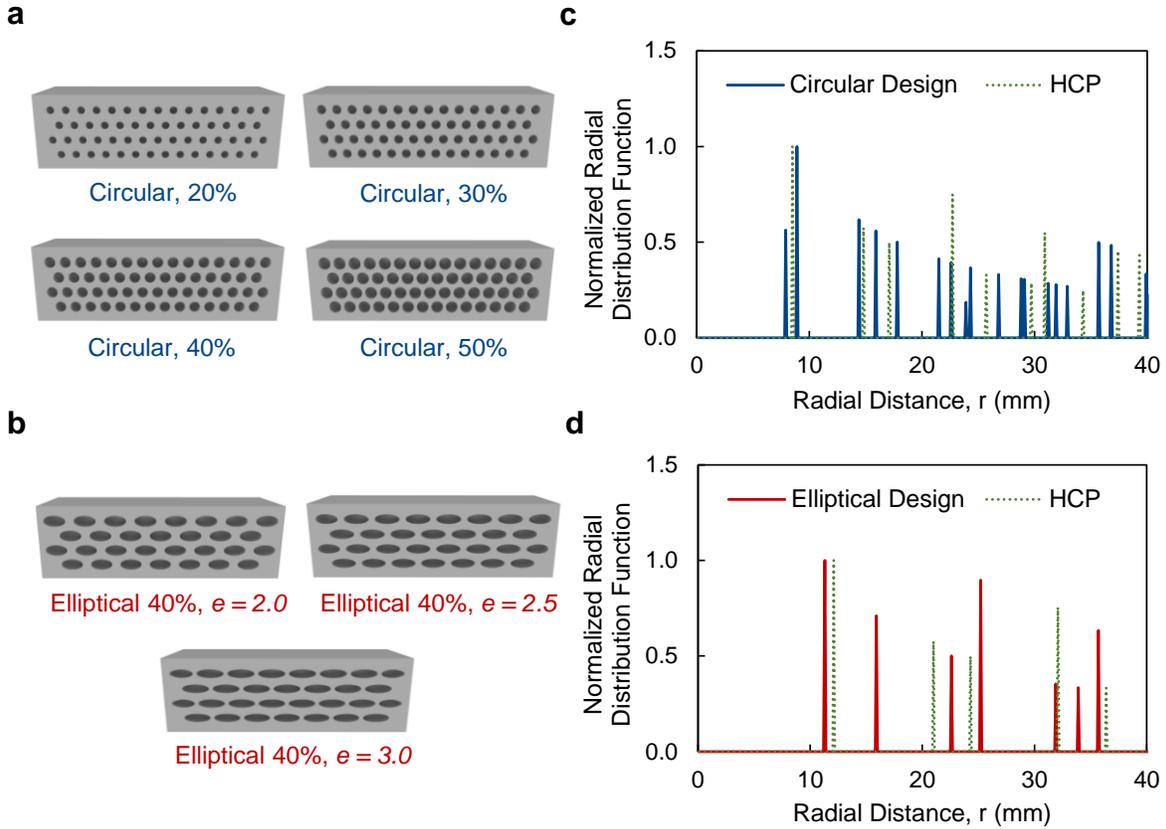

**Figure S2**. **Radial distribution function of circular and elliptical tubular designs**. a,b) Designs and corresponding c,d) normalized radial distribution function of circular and elliptical tubular architected materials.

The designs of circular and elliptical architected materials of different porosities ($\varphi = 20\%$, $30\%$, $40\%$, & $50\%$) and aspect ratios ($e = 2, 2.5, \& 3.0$) are presented in **Figure S2**a, b. The radial distribution functions, $g_2(r)$, are calculated for the (dissimilar) distribution of the center of tubes in the circular and elliptical architected materials. The normalized values of $g_2(r)$ with respect to peak value are presented for circular and elliptical architected materials in Figure S2c, d, respectively. In this study, the normalized $g_2(r)$ values for architected materials are benchmarked against those of the HCP lattice (Figure S2c, d). The findings reveal that, in terms of radial distance, the peak values of the normalized $g_2(r)$ for both circular and elliptical tubular materials do not align with the HCP lattice.

To capture the variation of the statistical parameter with degree of disorder, the translational ($T$) and orientational ($Q$) order parameters were determined for the range of 2D distributions of the particles, ranging from perfectly ordered, to disordered, and to random. In order to generate a broad range of disordered distribution of particles bound by random and perfectly ordered limits, the Metropolis algorithm for hard particles was employed.[18] In this algorithm, firstly, an initial configuration of the N particles is generated in a 2D bounding box such that no particles overlap one other. Then, a new configuration of particles is generated by moving a particle along each axis (x and y-axis) by a displacement, randomly distributed in the interval of $[-\delta, \delta]$, where $\delta$ is the maximum allowable 'step size' (the limit of the displacement). In the new configuration, if the displaced particle does not overlap with another particle, the move of the particle is accepted, otherwise, it is rejected. This process of moving the particle and accepting/rejecting the move is then repeated for all the N particles in the box to conclude one iteration. In our study, the initial configuration of the HCP lattice (perfectly ordered system) was chosen, and the degree of disorder was increased by increasing the *allowable* (maximum) step size ($\delta$) for displacing the particle.

As a result, choosing an HCP lattice for the initial configuration allows generating the particle distribution varying from ordered to disordered by increasing $\delta$ (**Figure S4a-d**). It should be further noted that the periodic boundary conditions were used to calculate the statistical parameters ($T$, $Q$, and $g_2(r)$), to use the small unit cell to approximate the large system of particles.

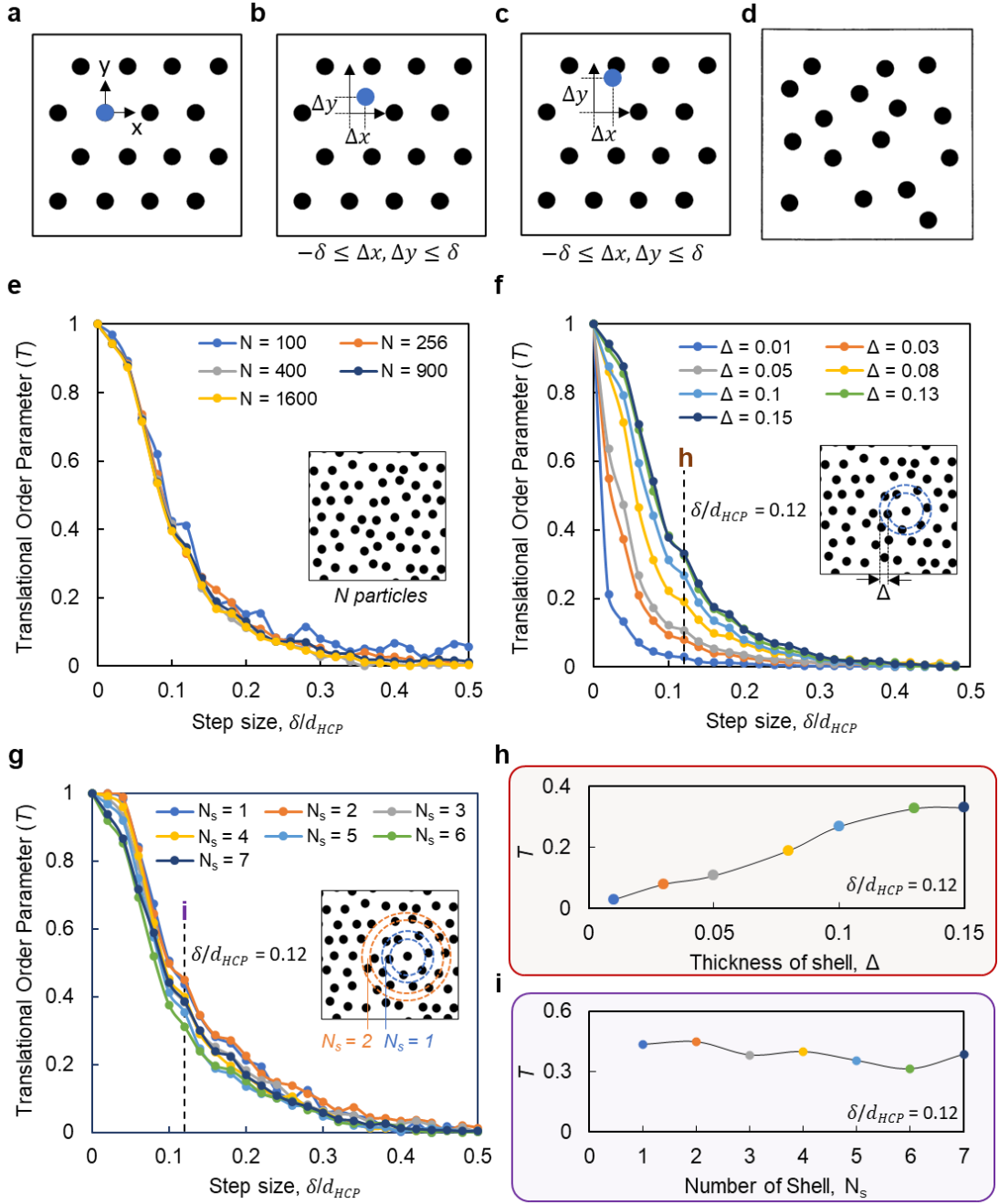

**Figure S3. Schematics of the Metropolis algorithm generates a disordered distribution** from a) an initial hexagonal close packing (HCP) lattice by moving a particle with an allowable step size $\delta$. A particle displacement is b) accepted if there is no overlap with another particle and c) rejected otherwise. d) The process repeats for all particles (one iteration) to generate a disordered distribution bound by HCP distribution and 2D random distribution. Parametric study of the translational order parameter ($T$) as a function of normalized step size, $\delta/d_{HCP}$ for varied e) number of particles or points, N, (f, h) thickness of the shell, $\Delta$, and (g, i) number of shells $N_s$.

The translational order parameter ($T$) depends on three variables, the number of particles in the unit cell (N), the thickness of the shell (Δ), and the number of shells ($N_s$).[20,21] Therefore, to calculate the $T$ parameter for particle systems with different degrees of disorder, a parametric study was performed with three variables (N, Δ, & $N_s$) as exhibited in **Figure S3**. It should be noted that $T$the parameter is 1 for the HCP lattice at $\delta = 0$. As $\delta$ approaches $0.5 d_{HCP}$ for all the different values of parameters, $T$ approaches 0, as shown in Figure S3a-c.

The increase in the number of particles in the unit cell (N) reduces the statistical noise in the $T$ parameter, as shown in Figure S3a, but rapidly increases the computational time. Therefore, considering the balance between the low statistical noise and the computational time, the number of particles in a unit cell was comfortably chosen as 900.

The shell size thickness, Δ, controls the occupation number, $n_i$, in a single shell used for the calculation of the $T$ parameter, as shown in Figure S1a-c. Increasing the Δ value higher than 0.15 can merge two (or more) surrounding peaks of $g_2(r)$ of the HCP lattice into one peak, meaning that the particles at two different distances in the HCP lattice will lie in the single shell. Therefore, 0.15 was kept as the upper limit of Δ. It can be noticed from Figure S3b, that the $T$ parameter increases with increasing Δ. Furthermore, at the fixed value of $\delta$, say $\delta = 0.12 d_{HCP}$, the $T$ parameter increases with $\delta$ at the decreasing rate and converges to 0.33 at the Δ = 0.15 as shown in Figure S3d. Considering the convergence, Δ = 0.15 was selected for the calculation of $T$ parameter. In the context of the number of shells, $N_s$, the $T$ parameter did not show any trend with the increasing $N_s$ as exhibited by Figure S3c. At the fixed value of $\delta$, say $\delta = 0.12 d_{HCP}$, it was found that $T$ parameter remained nearly constant. In this study, we have selected $N_s = 7$, based on the literature.[20]

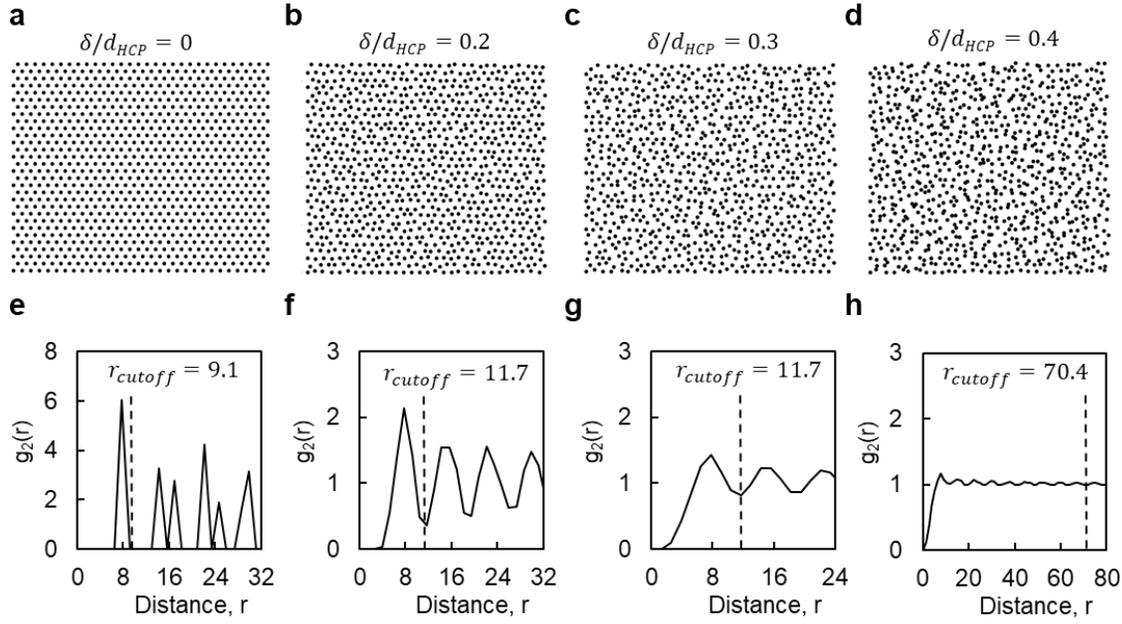

**Figure S4**. **Radial distribution function of perfectly ordered to highly disordered arrangements.** a-d) 2D – distribution of the 900 particles in a unit cell for maximum step size, $\delta$ varying from 0 to $0.4d_{HCP}$ and, e-h) the corresponding radial distribution function, $g_2(r)$ along with the cutoff radius, $r_{cutoff}$ (first minimum following the first peak in $g_2(r)$.

The orientational bond parameter ($Q_6$) depends on the total number of neighboring particles lying within the cutoff radial distance, $r_{cutoff}$, which is considered the first minimum following the first peak in $g_2(r)$.[20,21] As the degree of disorder in the system increases (with increasing $\delta/d_{HCP}$), the $r_{cutoff}$ increases, as shown by Figure S4e-h, It is worth noting that the $r_{cutoff}$ value increases by 7.7 times as the $\delta$ increases from 0 for the HCP lattice to $0.4d_{HCP}$ for a highly disordered system.

Considering the determined values from the parametric study, $T$ and $Q_6$ are determined for the tubular architected materials. Both the circular and elliptical architected material exhibit low translational order parameters, $T$, 0.21 and 0.30, respectively, which suggest that the periodic distribution of architecture has similar translational order parameters as particles in amorphous glasses.[20]

On the other hand, the circular architected materials exhibit an orientational order parameter, $Q_6$ = 0.77, which is higher than the $Q_6$ = 0.44 quantified for elliptical architected materials. The $Q$ of circular and elliptical architected materials are closer to the distribution of particles in crystal and amorphous glass, respectively.[20] The findings indicate for the first time, that for seemingly periodic design (distributions of features) in tubular architected materials, $Q$, more clearly represents the degree of order than $T$ for periodic distributions.

The use of the radial distribution function, along with the order parameters, alludes to the fact that although circular and elliptical tubular architecture appear periodic in terms of distribution, they are not perfectly ordered. This is counter-intuitive and indicates that periodicity is an insufficient measure to represent material distributions. Moreover, the spatial distribution of circular architected materials represents a higher orientational order parameter closer to a perfectly ordered arrangement (HCP) compared to the elliptical counterparts.

The findings can be examined and extended to other types of material distributions to understand the effectiveness of $Q$ vs. $T$ in capturing the degree of disorder in the field of architected materials. More importantly, the proper application of statistical mechanics and 'degree of disorder' can help quantify the degree of order and suggest a better measure than the qualitative assessment of 'periodicity'. In other words, the period vs. non-periodic conceptualization falls short of capturing the spectrum of degree of 'order' in the material (e.g., periodic distributions can be disordered whereas non-periodic distribution can have a very wide range of degrees of disorder which can determine the mechanical and wave propagation properties, none of which are captured by periodicity). In addition, the periodic vs. non-periodic classification is binary and lacks the capture of the spectrum that can be present in material arrangements.

The disordered designs in tubular geometries, orientations, and distributions can have the potential impact on material strength and fracture toughness. Our future work will focus on exploring a broader design domain encompassing variations in porosity, tube shape, and orientation, alongside a range of practical, manufacturable distributions. We aim to incorporate disorder parameters, $T$ and $Q$, as the key factors in characterizing these designs. Additionally, we plan to enhance our theoretical framework with crack propagation studies supported by numerical simulations, paving the way for innovative approaches to designing non-periodic and disordered systems in brittle materials. This initial study with the periodic distribution of the tubes serves as a foundational step towards employing circular and elliptical voids to optimize fracture properties, setting the stage for further investigation into more complex, polydisperse systems.

**Mechanical Response of Circular Architected Materials**
The effect of porosity on the fracture response and MOR of tubular architected materials with circular design is presented in **Figure S5**. The load-displacement plots obtained from the single-edge notch bend (SENB) test reveal distinct mechanical responses among various specimens (Figure S5a). The solid monolithic material, Circular, 20%, and Circular, 30% exhibit brittle failure. In contrast, Circular, 40% and Circular, 50% display stepwise cracking behavior characterized by multiple instances of hardening and softening. Notably, the crack-initiation load decreases with increasing porosity, $\varphi$, in tubular architected materials.

The R-curve, representing crack growth resistance (Δa) in terms of fracture toughness ($K_J$), also shows significant differences in fracture response among specimens (Figure S5b). The solid material, Circular, 20%, and Circular, 30% display nearly flat R-curves, indicating limited crack growth resistance. Conversely, Circular, 40% and Circular, 50% exhibit stepwise increases in the R-curve, signifying multiple steep rises in fracture toughness followed by sudden crack extensions. Circular, 40% demonstrates the highest growth in $K_J$, surpassing other designs and the solid counterpart.

A decreasing trend in average crack initiation fracture toughness ($K_{IC}$) is observed with increasing porosity, aligned with the reduced crack initiation load in the load-displacement plots. Circular, 20% and Circular, 30% exhibit $K_{IC}$ of $7.90 \pm 1.10\ MPa.mm^{0.5}$ and $6.84 \pm 1.08\ MPa.mm^{0.5}$, respectively, which are statistically similar to the solid material ($8.37 \pm 1.36\ MPa.mm^{0.5}$). However, Circular, 40% and Circular, 50% display $K_{IC}$ of $4.04 \pm 1.35\ MPa.mm^{0.5}$ and $3.22 \pm 0.85\ MPa.mm^{0.5}$ which are significantly lower than solid (Figure S5c). Nevertheless, no significant difference between all circular tubular material vs. solid material is found when specific values (normalized with respect to density) of $K_{IC}$ are used.

Intriguingly, despite decreasing $K_{IC}$ values with porosity, Circular, 40% and Circular, 50% demonstrate a substantial increase in average crack propagation fracture toughness ($K_{JC}$). The $K_{JC}$ of Circular, 40% and Circular, 50% are $18.89 \pm 6.02\ MPa.mm^{0.5}$ and $11.27 \pm 1.11\ MPa.mm^{0.5}$ which are significantly higher than the solid material by 2.25 and 1.34 folds. Circular, 20% and Circular, 30% demonstrate the $K_{JC}$ of $7.90 \pm 1.09\ MPa.mm^{0.5}$ and $9.74 \pm 1.85\ MPa.mm^{0.5}$ which are statistically similar to solid (Figure S5d).

The Circular, 20%, and Circular, 40%, exhibit modulus of rupture (MOR) of $3.16 \pm 0.36\ MPa$ and $3.41 \pm 0.68\ MPa$ which are statistically similar to that of the solid counterpart ($3.71 \pm 0.38\ MPa$). However, Circular, 30% and Circular, 50% demonstrate MOR of $2.70 \pm 0.22\ MPa$ and $2.61 \pm 0.50\ MPa$ which are significantly lower than the solid material. The average MOR of tubular architected materials presents an expected decreasing trend with increasing porosity, except for Circular, 40% (Figure S5e). An F-test and a T-test with a confidence level of 95% are used for all statistical analyses.

The underlying response that generates the trend in fracture toughness is further elaborated in this Supporting Information (Theoretical and Experimental Analysis Section) by discussing the theoretical stress intensity factors ($K_I$) in simplified analyses under far-field tension.

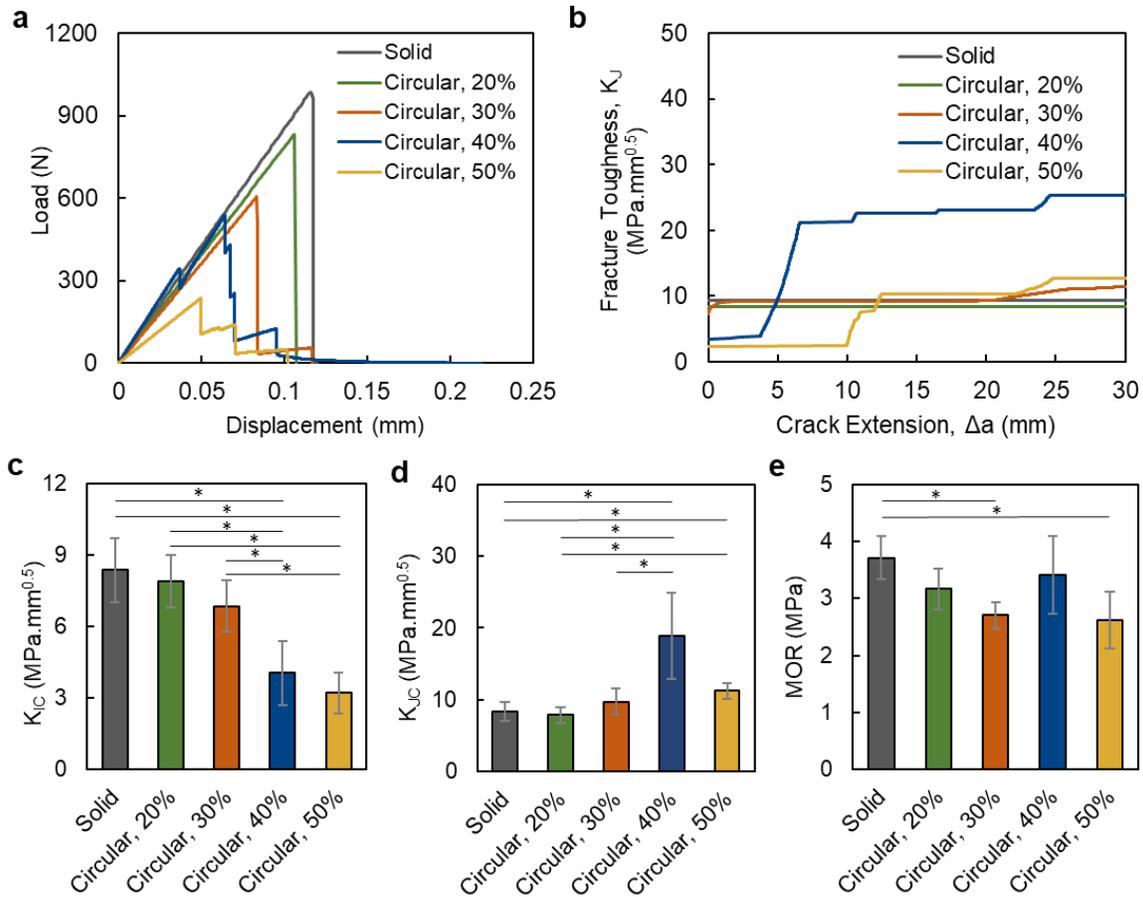

**Figure S5**. **Mechanical response of tubular architected cement-based material with circular design compared to monolithic solid.** a) Load – displacement plots obtained from SENB specimens, b) R-curve demonstrating the toughness vs. crack extension, c, d, and e) Crack initiation fracture toughness ($K_{IC}$), fracture toughness ($K_{JC}$), and modulus of rupture (MOR) of the architected material compared to monolithic solid. Data is shown as mean ± SD. '∗' depicts $p < 0.05$ which indicates the statistically significant difference between the samples (at the ends of the solid line below a '∗'). *p*-value is obtained from the F-test and T-test.

**Mechanical Response of Elliptical Architected Materials**

Figure S5 establishes that the tubular architected material with 40% porosity displays the highest fracture toughness compared to other circular designs. Hence, a porosity of 40% is employed to study the effect of the aspect ratio of tubes on the mechanical response of the tubular architected material. The aspect ratio, indicating the ratio of the semi-major axis to the semi-minor axis, is varied from 2 to 3 ($e = 2, 2.5, \& 3$) in this study.

The L-D plot in **Figure S6**a illustrates that solid material, Elliptical, 40% $e = 2$, and Elliptical, 40% $e = 3$ fail nearly in a brittle fashion whereas Elliptical, 40% $e = 2, e = 2.5, e = 3$ demonstrate stepwise failure with $e = 2.5$ outperforming other the cases. Consequently, Elliptical, 40% $e = 2.5$ demonstrates a notable rise in the R-curve compared to other elliptical as well as the solid materials (Figure S6b).

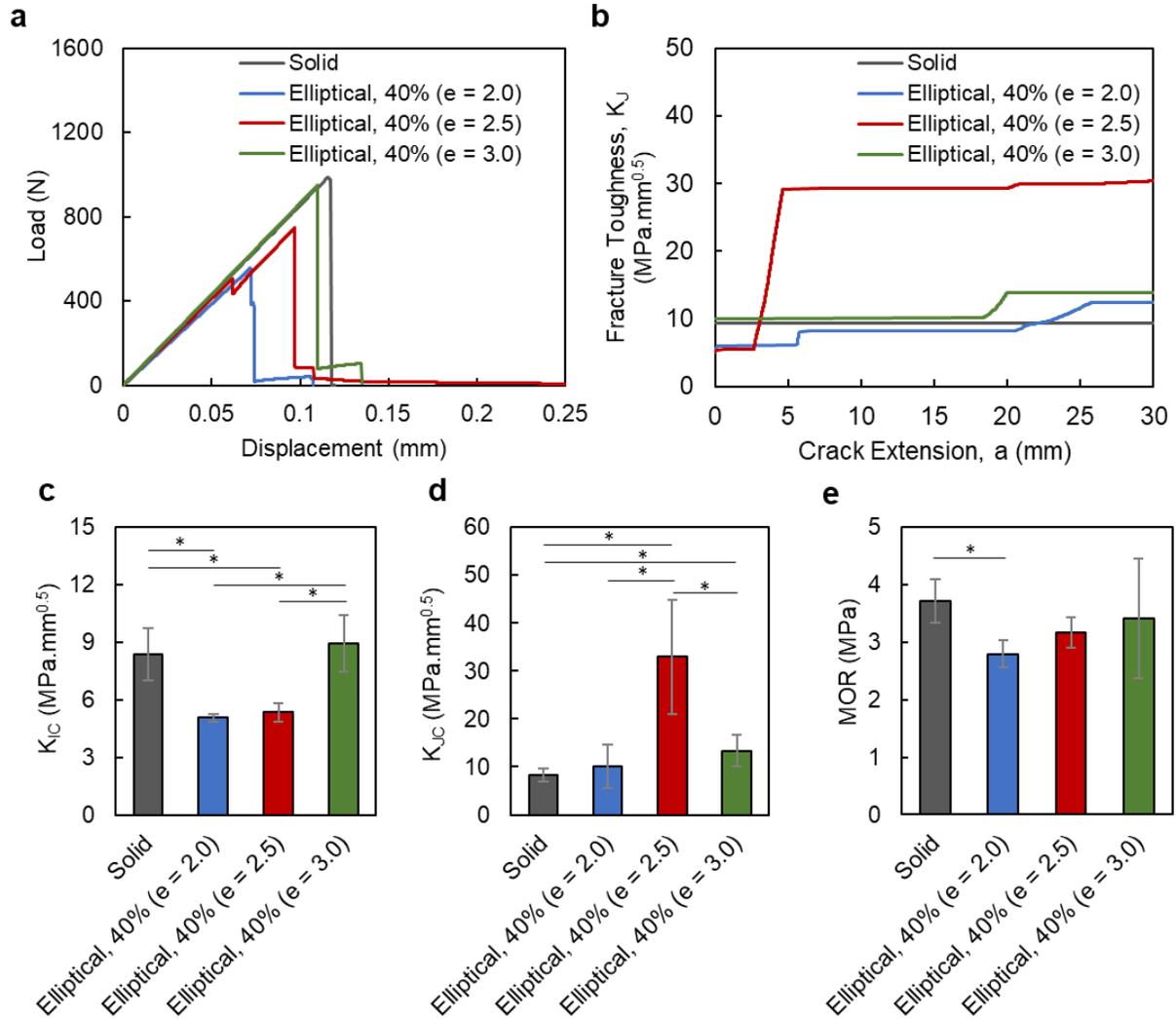

**Figure S6**. **Mechanical response of tubular architected cement-based material with elliptical design compared to monolithic conventionally cast solid.** a) Load – displacement plots obtained from SENB specimens, b) R-curve showing the toughness vs. crack extension, c, d, and e) Crack initiation fracture toughness ($K_{IC}$), fracture toughness ($K_{JC}$), and modulus of rupture (MOR) of the architected material compared to monolithic cast. Data is shown as mean ± SD. '∗' depicts $p < 0.05$ which indicates the statistically significant difference between the samples (at the ends of the solid line below a '∗'). $p$-value is obtained from the F-test and T-test.

Examining Figure S6c, we observe an increasing trend in average $K_{IC}$ with aspect ratio, $e$, in tubular architected material. The $K_{IC}$ of Elliptical, 40%, $e = 3$ is $8.94 \pm 1.47\ MPa.mm^{0.5}$ which is statistically similar to the solid material. However, Elliptical, 40% $e = 2$ and Elliptical, 40% $e = 2.5$ exhibits $K_{IC}$ of $5.05 \pm 0.21\ MPa.mm^{0.5}$ and $5.36 \pm 0.48\ MPa.mm^{0.5}$ which are statistically lower compared to solid material. On the other hand, the Elliptical, 40% $e = 2.5$ demonstrates $K_{JC}$ of $33.24 \pm 11.90\ MPa.mm^{0.5}$ which is significantly higher compared to solid and other elliptical designs. However, elliptical-designed tubular materials with $e = 2.0$ and $e = 3.0$ demonstrate $K_{JC}$ of $10.09 \pm 4.50\ MPa.mm^{0.5}$ and $13.34 \pm 3.28\ MPa.mm^{0.5}$ compared to solid case (Figure S6d).

Additionally, the Elliptical, 40% $e = 2.5$ and $e = 3$ exhibit MOR of $3.17 \pm 0.26\ MPa$ and $3.41 \pm 1.03\ MPa$ which are statistically similar to solid, while the Elliptical, 40% $e = 2$ displays MOR of $2.80 \pm 0.23\ MPa$ which is significantly lower compared to the solid material (Figure S6e). An F-test and a T-test with a confidence level of 95% are used for all statistical analyses.

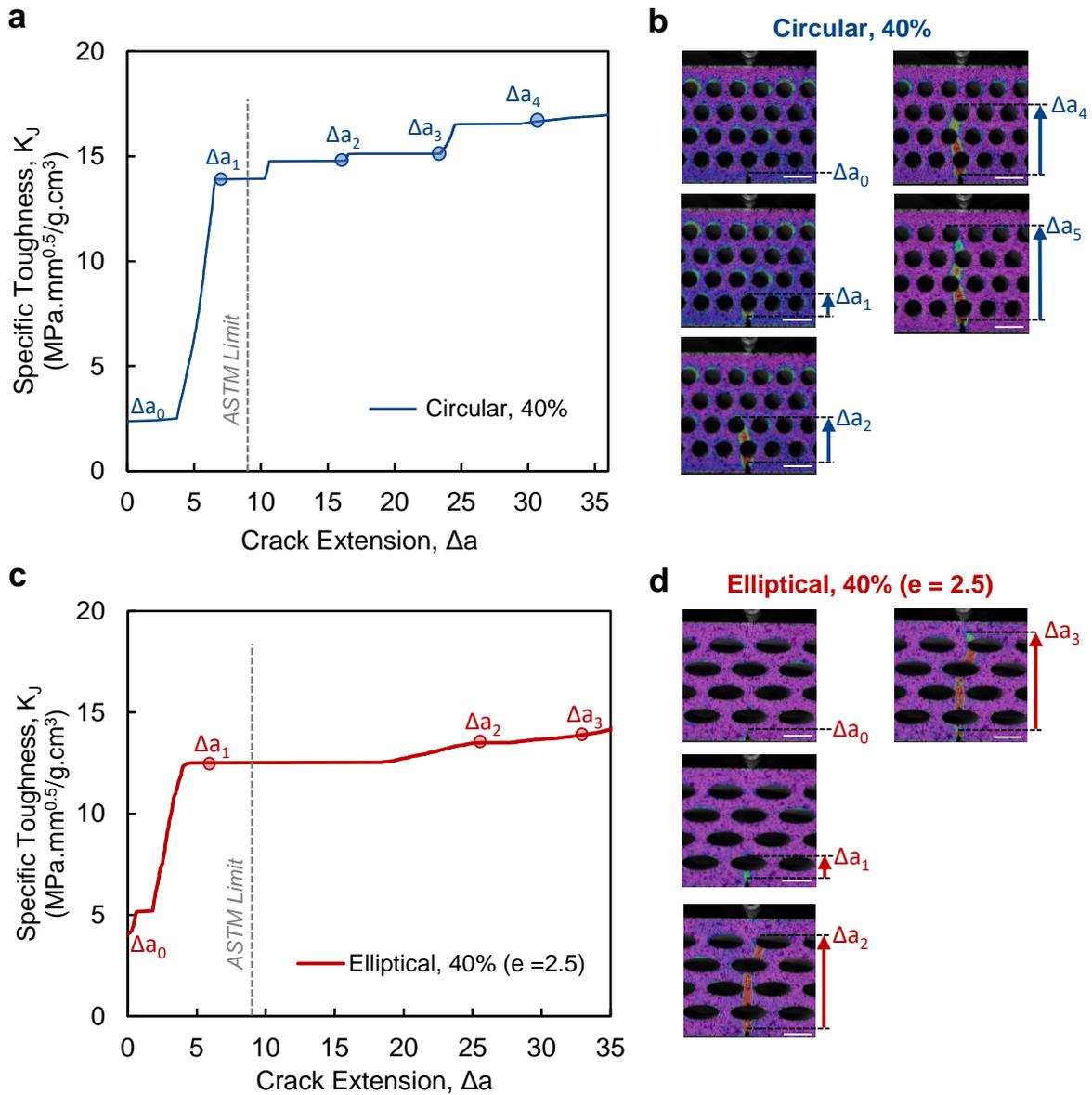

**Figure S7**. a,c) R-curve representing the specific fracture toughness vs. crack extension (shows as a line) of Circular, 40% and Elliptical, 40% marked with b,d) crack extension ($\Delta a_i$) obtained from DIC analysis during fracture experiments (shown as circle points).

The crack extension, determined through load-line displacement and calculated using the compliance method, can be corroborated for further validation of the rising R-curve and accuracy of the crack extension, by comparing it with the crack extension obtained through Digital Image Correlation (DIC) analysis during fracture experiments. The crack extensions, $\Delta a_i$, denoted on the R-curve in **Figure S7**a and c, correspond to the stepwise crack progression from tube in one layer to the another in both Circular, 40% and Elliptical, 40%, from the DIC, as shown in Figure S7b and d, respectively. These $\Delta a_i$ are positioned in close proximity to the onset of each step of the crack growth in the X-axis of the R-curve, indicating that the calculated crack extension is well-aligned with the experimentally observed crack extensions from DIC analysis initiated at the tubes and extended into the subsequent layers.

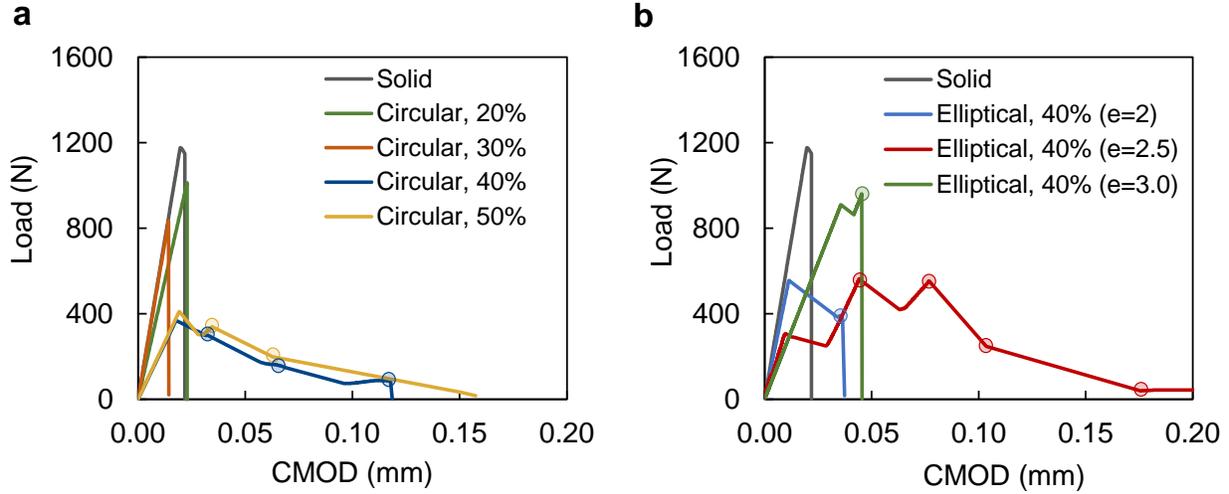

**Figure S8.** Load vs. Crack mouth opening displacement (CMOD) plots of tubular architected materials with a) circular and b) elliptical designs. Circle marks indicate the crack initiation from the tube.

The stepwise crack progression in tubular architected materials is also shown in circular and elliptical designs using load vs. crack mouth opening displacement (CMOD) plots obtained from the SENB test in **Figure S8**a and b, respectively. The solid material and tubular architected material with a lower range of porosity ($\varphi$ = 20% and 30%) demonstrate brittle failure. The elliptical designs with $e$ = 2.0 and 3.0 exhibit single CMOD extension followed by a delayed brittle failure. On the other hand, the multiple extensions of CMOD associated with the extension of the crack along the depth are observed in Circular, 40%, Circular, 50%, and Elliptical, 40% $e$ = 2.5. The crack propagation throughout the depth of the sample and increased CMOD takes place at the initiation of the crack from the tube as marked by circles.

The underlying mechanism for the effect of the geometric attributes of the tube (circular vs. elliptical) on the fracture toughness is further elaborated in this Supporting Information (Effect of the Shape and Location Section) by analyzing the theoretical stress intensity factor under far-field tension.

**Theoretical and Experimental Analysis of Stress Intensity Factor Due to the Presence of Tube Ahead of a Notch in Tubular Architected Materials**

The stress intensity factor characterizes the intensity of the stress field in a small region surrounding the crack tip.[23] The stress intensity factor, $K_{I,Solid}$, of the crack tip in the monolithic elastic solid material under the far-field tensile stresses, $\sigma_{x\infty}$ (as shown in **Figure S9**a), is defined using the Eq. S15.[24]

$$K_{I,Solid} = \sigma_{x\infty}\sqrt{\frac{\pi c}{2}} \quad (Eq.S15)$$

where c is the pre-existing crack (notch) length. The presence of tubes in the material alter the stress intensity factor, $K_{I,Tubular}$, of the notch tip.[24,25] The effect of a single tube on the stress intensity factor of a crack tip (Figure S9b) is the first step to understand crack-tube interaction and is determined by using the stress-based approach in this study.[24,25] The first order solution of stress intensity factor due to the presence of a single tube, $K_{I,Tubular}$, is determined using the Eq. S16.[24]

$$K_{I,Tubular} = \sqrt{\frac{2}{\pi c}} \int_{y_1}^{y_2} \sigma_x(x,y) \left(\frac{y-y_2}{y_1-y}\right)^{0.5} dy \quad (Eq.S16)$$

where, $\sigma_x(x,y)$ is the stress on the pre-existing crack (notch) line in the uncracked body due to the hollow tube as illustrated in Figure S10b. $y_1$ and $y_2$ are the relative position of the closest and farthest crack tip with respect to the center of the tube, respectively, as shown in Figure S9b.

The $\sigma_x(x,y)$ is further determined under uniaxial far-field tension using the Eq. S17.[26]

$$\sigma_x(x,y) = \frac{\alpha + \bar{\alpha}}{2} \quad (Eq.S17)$$

where,

$$\alpha = \frac{\sigma_{x\infty}}{4}\Big[\{(A+\bar{A}) + (B\coth(\zeta) + \bar{B}\coth(\bar{\zeta}))\}$$
$$- \frac{1}{\sinh^3(\zeta)}\{-(B\cosh(\bar{\zeta}) + C\cos(\zeta)) + 4D\sinh^3(\zeta) + 2E\cosh(\zeta)(2\sinh^2(\zeta)$$
$$- 1)\}\Big] \quad (Eq.S18)$$

and $\zeta$ is the coordinate in the transformed plane and it is the function of semi-major and semi-minor lengths $(a,b)$ of the elliptical tube with the center of tube at (0,0). $\zeta$ is defined by Eq. S19.[26]

$$\zeta = \cosh^{-1}\left[\frac{x+iy}{\sqrt{a^2-b^2}}\right] \quad (Eq.S19)$$

On the other hand, A, B, C, D, and E are the geometrical parameters determined by Eq. S20-S24.[26]

$$A = e^{2\xi_0} \quad (Eq.\,S20)$$

$$B = 1 - e^{2\xi_0} \quad (Eq.\,S21)$$

$$C = 1 - \cosh(2\xi_0) \quad (Eq.\,S22)$$

$$D = -\frac{1}{2} e^{2\xi_0} \cosh(2\xi_0) \quad (Eq.\,S23)$$

$$E = -\frac{1}{2} e^{2\xi_0} \sinh(2\xi_0) \quad (Eq.\,S24)$$

where $\xi_0$ represents the tube boundary and can be defined by Eq. S25.[26]

$$\xi_0 = \cosh^{-1}\left[\frac{a}{\sqrt{a^2 - b^2}}\right] \quad (Eq.\,S25)$$

The value of $K_{I,Tubular}$ obtained from Eq. S16 is validated by comparing them to values of $K_I$ provided by Tirosh and Tetelman, 1976,[24] for circular tubes of different sizes used in this work.

Furthermore, the stress intensity factor ratio, $F(y_1, y_2)$, can be defined as the ratio of $K_I$ of tubular architected material to that of solid and is defined by Eq. S26.

$$F(y_1, y_2) = \frac{K_{I,Tubular}}{K_{I,Solid}} = \frac{1}{\sigma_{x\infty}} \frac{2}{\pi c} \int_{y_1}^{y_2} \sigma_x(x,y) \left(\frac{y - y_2}{y_1 - y}\right)^{0.5} dy \quad (Eq.\,S26)$$

The stress intensity factors, $K_I$, for solid, Circular, 40%, and Elliptical, 40% $e = 2.5$, as representative tubular cases, determined using Eq. S16. The results are plotted for $K_I$ against the far-field stress, $\sigma_{x\infty}$, in Figure S9c and demonstrate the linear increase in $K_I$ with the increase in far-field stress, $\sigma_{x\infty}$. However, according to Griffth's energy balance approach, the condition for crack initiation is for the stress intensity factor ($K_I$) to reach the critical value or the fracture toughness ($K_{IC}$).[27]

$K_{IC}$ of the cement paste obtained from the SENB experiments is plotted in Figure S10c (as shown by the gray line). The intersection of $K_I$ vs $\sigma_{x\infty}$ plot with the critical stress intensity factor, $K_{IC}$, yields the far-field stress at failure, $\sigma_{x\infty,F}$, as illustrated in Figure S9c for solid, and tubular Circular, 40%, and Elliptical, 40% $e = 2.5$ cases. It is found that for a single tube ahead of the crack tip, the far-field stresses at failure for tubular, $\sigma_{x\infty,F}^{Tubular\,(circular\,or\,elipitical)}$, are both smaller than that of a monolithic solid, $\sigma_{x\infty,F}^{Solid}$, as shown in Figure S9c.[24] This indicates that the required stress for initiation of the crack from the notch tip is lower in the case where tubes are present compared to the solid case without any tube ahead of the tip. Thus, $F(y_1, y_2) > 1$ when a tube is present ahead of the crack tip.

Therefore, by equating the Eq. S15 and Eq. S16 to $K_{IC}$ and using Eq. S26, the relation between the far-field stresses at failure for tubular material, $\sigma_{x\infty,F}^{Tubular\ (circular\ or\ elipitical)}$, and solid, $\sigma_{x\infty,F}^{Solid}$, can be written as Eq. S27.

$$\sigma_{x\infty,F}^{Tubular} = \frac{\sigma_{x\infty,F}^{Solid}}{F(y_1, y_2)} \quad (Eq.S27)$$

Assuming that the far-field stresses are directly proportional to the applied load in linear elastic material, the load at failure for tubular material, $P_F^{Tubular}$, and solid, $P_F^{Solid}$, can be illustrated by Eq. S28.

$$P_F^{Tubular} = \frac{P_F^{Solid}}{F(y_1, y_2)} \quad (Eq.S28)$$

And the normalized load at failure, $P_{F,N}$, of tubular architected material with respect to solid can be defined using the Eq. S29.

$$P_{F,N} = \frac{P_F^{Tubular}}{P_F^{Solid}} = \frac{1}{F(y_1, y_2)} \quad (Eq.S29)$$

The stress intensity factor ratio, $F(y_1, y_2) = K_{I,Tubular}/K_{I,Solid}$, for different tubes is calculated using Eq. S26 and the results are presented in Figure S9d. It is found that $F(y_1, y_2)$ increases with the increasing porosity (diameter) in the circular tube and decreases with the increasing aspect ratio in elliptical tubes. This indicates that the stress intensity factor can increase by as high as a factor of 1.5 due to the presence of a tube in front of the crack tip (Figure S9d).

Therefore, based on Eq. S29, the normalized load at failure, $P_{F,N}$, follows the opposite trend where it decreases with the increasing porosity (diameter) of the tube and increases with increasing aspect ratio as illustrated in Figure S9e. This theoretical trend is intuitive and indicates that the peak load decreases by a factor of 0.75 due to the addition/presence of the tube.

These trends of the theoretical $P_{F,N}$, in relation to porosity and aspect ratio, correspond closely with the trends of experimental $P_{F,N}$ derived from the SENB test on both tubular architected and solid materials, as shown in Figure S9f. However, there is a noticeable difference between the experimental and theoretical values of $P_{F,N}$ and the theoretical at the lower bounds which has to do with the superposition of other tubes in the experiments. In the theoretical analysis of the stress intensity factor, only the effect of a single tube on the stress intensity factor is considered whereas experimentally the stress intensity factor at the crack tip is also influenced by the presence of additional tubes in the architected material.

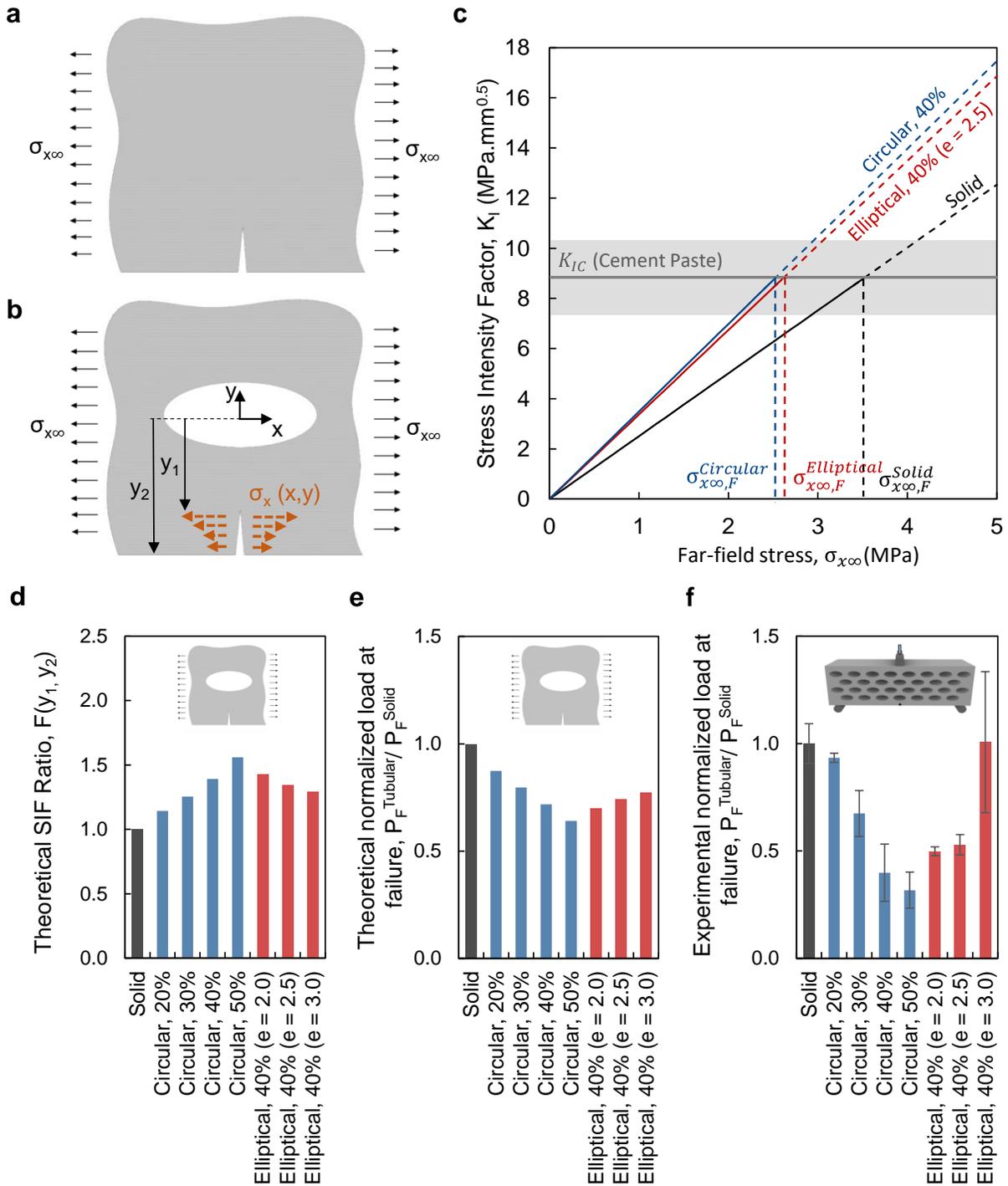

**Figure. S9. Theoretical and experimental analysis of stress intensity factor.** a) Schematics of a pre-existing crack (notch) in the monolithic material and b) the interaction of hollow tube with the crack under far-field tensile stresses, c) Stress intensity factor ($K_I$) vs. far-field stress at failure for solid, Circular, 40%, and Elliptical, 40% $e = 2.5$ along with the critical stress intensity

factor of cement paste, d) Theoretical Mode-I stress intensity factor (SIF) ratio of tubular materials vs. solid counterpart obtained from Eq. S26, e) Theoretical and f) experimental normalized load at failure of architected materials vs. solid counterpart.

**Effect of the Shape and Location of the Tube on Mode-I Stress Intensity Factor in Tubular Architected Materials**

In considering the superposition of the tube, the location and the shape of the tube relative to the orientation of the crack (notch) presents additional alteration to the stress intensity factor at the notch tip. **Figure S10** illustrates how the normalized Mode-I stress intensity factor, $F(y_1, y_2)$, varies depending on the tube's location along the x-axis, $x_o$, in relation to the crack tip for Circular, 40% and Elliptical, 40% $e = 2.5$.

The influence of the circular or elliptical tubes on the Mode-I stress intensity factor ratio, $F(y_1, y_2)$, of the crack tip is most pronounced when the tube is directly in front of the crack tip, specifically when $x_o = 0$ (Figure S9). As the tube begins to move away from the crack tip, as indicated by an increase in $x_o$, the stress intensity factor ratio rapidly decreases. When the circular or elliptical center point of the tube is located at $x_o$ of 7.0 mm or 9.0 mm respectively, the stress intensity factor ratio is equal to 1, indicating no effect on the stress intensity factor ratio.

Furthermore, as the distance of the tube from the notch continues to increase, the Mode-I stress intensity factor ratio drops below 1. Considering the inverse relation between load at failure and stress intensity factor ratio based on Eq. S29, the presence of a tube at a distance (larger than 7.0 and 9.0 mm) amplifies the load required for failure at the crack tip. However, at greater distances (larger than ~ 60 mm), the normalized stress intensity factor reverts to approach the value of 1, which signifies the diminishing effect of the presence of a tube on the stress intensity factor at the notch. Therefore, if the superposition of the tubes is to be considered, the competing effects will be in play on the required load at failure, especially between the tubes within or beyond the stress intensity factor ratio of 1. The theoretical determination of the stress intensity factor can be extended for comparison with experiments considering both superimposing the effect of multiple tubes at various locations ahead of the crack as well as the bending stress state.

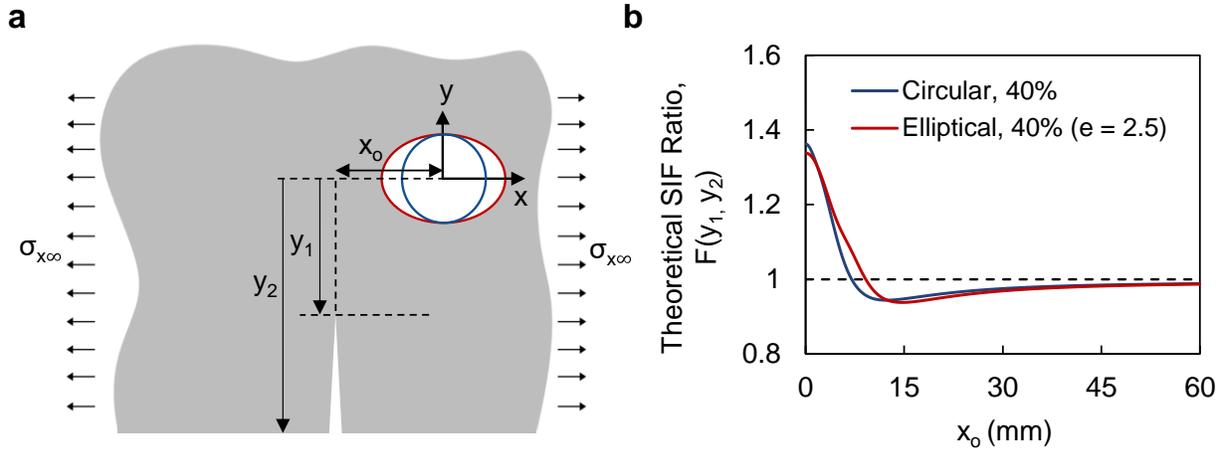

**Figure S10.** a) A single tube located at the offset of $x_o$ with respect to crack tip, b) Theoretical stress intensity factor ratio, $F(y_1, y_2)$, of notch tip with respect to offset, $x_o$, for circular and elliptical tube obtained from Eq. S26.

It is important to note that the increase in the aspect ratio of the tube (ellipticity) leads to decrease in the Mode-I stress intensity factor, which in turn increases the load-at-failure with increasing aspect ratio (*e*) as shown in Figure S9 d-f and **Figure S11a, c**. The increase in load-at-failure subsequently lead to higher resistance to crack initiation, hence, resulting in an increase in the crack initiation fracture toughness, $K_{IC}$, of the tubular architected materials with increasing aspect ratio.

The orientation of the tube can principle also alter the stress intensity factor, $K_{IC}$. Positioning elliptical tube along their longer axis, as shown in Figure S11b, which involves rotating the tubes by 90 degrees, orients the more highly curved end of the elliptical tube toward the crack. This orientation increases the mode-I theoretical stress intensity factor (SIF) by 16 – 20% compared to when elliptical tube is aligned along the shorter axis, as illustrated in Figure S11c. This increase in stress intensity factor reduces the load-at-failure and, consequently, lowers the crack initiation fracture toughness ($K_{IC}$) when tubes are oriented along their longer axis.

Additionally, once the crack is propagation from the notch to the elliptical tube, a higher stress concentration factor is developed in the elliptical tubes that is aligned along the longer axis compared to those aligned along the shorter axis.[26,28] In that case, in context of step-wise cracking for crack reinitiation from the tube (and subsequent tubes), we can hypothesize that higher stress concentration in "along the longer axis" case, leads to a lower load-at-failure compared to the "along the shorter axis" case, thus could negatively impact the stepwise cracking mechanism. In other words, a circular design is hypothesized to outperform the "along longer axis" design and underperform the "along the longer axis" design in terms of step-wise cracking from the tube.

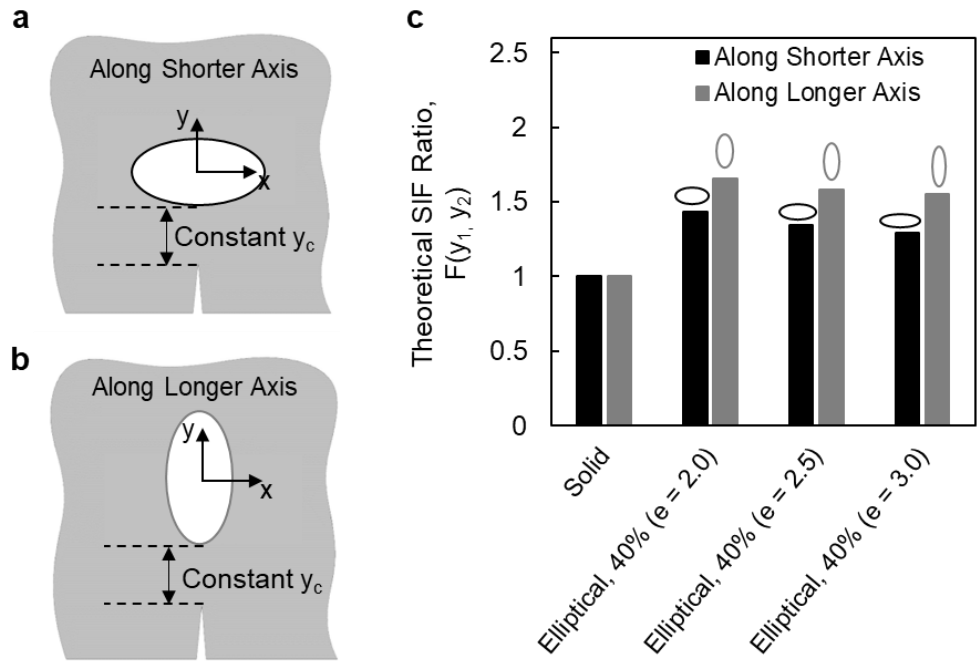

**Figure S11. A single elliptical tube positioned ahead of the crack in two orientations, leading to variation on theoretical stress intensity factor**: a) with the shorter axis aligned with the crack and b) with the longer axis aligned with the crack. In both scenarios, the distance between the crack tip and the front of the tube, labeled as $y_c$, is maintained constant for a specific aspect ratio, c) Theoretical stress intensity factor ratio, $F(y_1, y_2)$, of the notch tip for elliptical tubes with different aspect ratios for two different orientations.

**Movies**.

Movie S1. Digital Image Correlation (DIC) of the SENB testing of Circular architected materials with 40% porosity.

Link: https://youtu.be/qIQi26O5RP8

Movie S2. Digital Image Correlation (DIC) of the SENB testing of Elliptical architected materials with 40% porosity and aspect ratio, e = 2.5.

Link: https://youtu.be/JlM-xXBLfUE